\documentclass[acmsmall]{acmart}
\usepackage{xspace}
\newcommand{\tool}{\textsc{PtoP}\xspace}

\usepackage{multirow}
\usepackage{amssymb}
\usepackage{color}
\usepackage{subcaption}
\usepackage{algorithm,algorithmic}
\usepackage{amsmath}
\usepackage{multicol}
\usepackage{hyperref}
\usepackage{url}
\usepackage{enumitem}
\definecolor{todocolor}{rgb}{0.9,0.1,0.1}

\setcopyright{cc}
\setcctype{by}
\acmDOI{10.1145/3808153}
\acmYear{2026}
\acmJournal{PACMSE}
\acmVolume{3}
\acmNumber{FSE}
\acmArticle{FSE146}
\acmMonth{7}
\acmSubmissionID{fse26mainb-p753-p}
\received{2025-09-12}
\received[accepted]{2026-03-24}

\begin{document}

%%
%% The "title" command has an optional parameter,
%% allowing the author to define a "short title" to be used in page headers.
%%
% \title{SVGD-Guided Adversarial Traffic for Hazardous Scenario Generation}

% \title{Driving Diversity: SVGD-Based Hazardous Scenario Synthesis for ADS Testing}
%\title{MART: Towards Evaluating the Robustness of Autonomous Driving Systems in Multi-Vehicle Scenario Online}

%%
%% The "author" command and its associated commands are used to define
%% the authors and their affiliations.
%% Of note is the shared affiliation of the first two authors, and the
%% "authornote" and "authornotemark" commands
%% used to denote shared contribution to the research.

\title{From Particles to Perils: SVGD-Based Hazardous Scenario Generation for Autonomous Driving Systems Testing}

\author{Linfeng Liang}
\orcid{0000-0001-5442-3656}
\affiliation{%
  \institution{Macquarie University}
  \city{}
  \country{Australia}
}
\email{linfeng.liang@hdr.mq.edu.au}

\author{Xiao Cheng}
\orcid{0000-0001-5456-3827}
\affiliation{%
  \institution{Macquarie University}
  \city{}
  \country{Australia}
}
\email{xiao.cheng@mq.edu.au}

\author{Tsong Yueh Chen}
\orcid{0000-0003-3578-0994}
\affiliation{%
  \institution{Swinburne University of Technology}
  \city{}
  \country{Australia}
}
\email{tychen@swin.edu.au}

\author{Xi Zheng}
\authornote{Corresponding author}
\orcid{0000-0002-2572-2355}
\affiliation{%
  \institution{Macquarie University}
  \city{}
  \country{Australia}
}
\email{james.zheng@mq.edu.au}
% \author{John Smith}
% \affiliation{%
%   \institution{The Th{\o}rv{\"a}ld Group}
%   \city{Hekla}
%   \country{Iceland}}
% \email{jsmith@affiliation.org}

% \author{Julius P. Kumquat}
% \affiliation{%
%   \institution{The Kumquat Consortium}
%   \city{New York}
%   \country{USA}}
% \email{jpkumquat@consortium.net}

%%
%% By default, the full list of authors will be used in the page
%% headers. Often, this list is too long, and will overlap
%% other information printed in the page headers. This command allows
%% the author to define a more concise list
%% of authors' names for this purpose.
% \renewcommand{\shortauthors}{Trovato et al.}

%%
%% The abstract is a short summary of the work to be presented in the
%% article.

\begin{abstract}
Simulation-based testing of autonomous driving systems (ADS) must uncover realistic and diverse failures in dense, heterogeneous traffic. However, existing search-based seeding methods (e.g., genetic algorithms) struggle in high-dimensional spaces, often collapsing to limited modes and missing many failure scenarios. We present \tool, a framework that combines adaptive random seed generation with Stein Variational Gradient Descent (SVGD) to produce diverse, failure-inducing initial conditions. SVGD balances attraction toward high-risk regions and repulsion among particles, yielding risk-seeking yet well-distributed seeds across multiple failure modes. \tool\ is plug-and-play and enhances existing online testing methods (e.g., reinforcement learning--based testers) by providing principled seeds. Evaluation in CARLA on two industry-grade ADS (Apollo, Autoware) and a native end-to-end system shows that \tool\ improves safety violation rate (up to 27.68\%), scenario diversity (9.6\%), and map coverage (16.78\%) over baselines.

\end{abstract}

%%
%% The code below is generated by the tool at http://dl.acm.org/ccs.cfm.
%% Please copy and paste the code instead of the example below.
%%


%%
%% Keywords. The author(s) should pick words that accurately describe
%% the work being presented. Separate the keywords with commas.
%% A "teaser" image appears between the author and affiliation
%% information and the body of the document, and typically spans the
%% page.

% \received{20 February 2007}
% \received[revised]{12 March 2009}
% \received[accepted]{5 June 2009}
% \setcopyright{none} % to remove the copyright notice
% \settopmatter{printacmref=false} % to remove the ACM Reference Format
% \renewcommand\footnotetextcopyrightpermission[1]{}
%%
%% This command processes the author and affiliation and title
%% information and builds the first part of the formatted document.
\maketitle

\section{Introduction}

Autonomous driving systems (ADS)~\cite{ap,tesla_autopilot,chen2024end} are safety-critical cyber-physical systems that require rigorous testing to ensure reliability prior to deployment~\cite{bertoncello2015ten,barr2014oracle}. While real-world testing provides valuable insights, it is costly and time-consuming~\cite{lambert2016understanding}, making simulation-based testing~\cite{deng2022scenario,liang2023rlaga,cheng2024decictor,chen2024misconfiguration} a practical alternative for evaluating ADS across diverse scenarios. To be effective, however, simulation must bridge the gap to real-world conditions by moving beyond simplified assumptions and uncovering safety violations in realistic settings with \emph{dense traffic and complex interactions among heterogeneous dynamic objects} (e.g., vehicles, cyclists, and pedestrians)~\cite{attewell1994crashes,RoadSafetyDataHub2025}.

%Existing ADS testing methods typically follow a two-stage process\jz{this is not correct statement.} 
Instead of generating pre-defined trajectories for dynamic objects---which is often unrealistic and exacerbates the simulation-to-reality gap---some latest approaches \cite{liang2023rlaga} adopt a two-stage process 
(1) \textit{Offline seeding}---generating initial states of dynamic objects (e.g., spawn positions, 
velocities, and orientations of vehicles, cyclists, and pedestrians) relative to the ego vehicle; and 
(2) \textit{Online testing}---the dynamic update of object trajectories based on complex interactions with other participants in the scenario, including the ego vehicle, governed by predefined policies or traffic models.
These updates are commonly achieved through approaches such as reinforcement learning (RL)-based methods~\cite{liang2023rlaga} or gradient-based methods~\cite{hanselmann2022king}.
Nevertheless, the quality of the offline seeding stage remains highly crucial: small
perturbations to initial conditions can yield substantially different scenario evolutions and thus
alter the likelihood of failures.

Seeding is commonly cast as a search problem and addressed with search-based techniques such as genetic algorithms (GAs)~\cite{tian2022mosat,li2020av,liang2023rlaga,cheng2024decictor,feng2023dense}. 
However, under dense heterogeneous traffic, such methods face two challenges: \textit{effectiveness}---identifying rare, failure-inducing configurations; and \textit{diversity}---covering a wide range of distinct failure modes. 
These stem from the \emph{high-dimensional search space} induced by multiple object types, where failure cases cluster in disjoint regions. Classical GA operators often converge prematurely and struggle to preserve diversity, leading to violations concentrated in a few clusters. 
For example, in the TOWN04 map of CARLA~\cite{dosovitskiy2017carla}, 372 spawn points with just five surrounding vehicles yield $2.545\times 10^{15}$ possible initial configurations. Adding continuous parameters (e.g., headings, speeds) and other agents like pedestrians further expands the space, making exhaustive or unguided search infeasible.

To address the seeding challenges while remaining agnostic to online testing, we present \tool, a general-purpose ADS testing framework that generates diverse initial conditions while seamlessly integrating with existing online testing tools~\cite{liang2023rlaga,hanselmann2022king} to enhance the overall performance.
A seed pool is maintained by an adaptive random seed generator (ARSG), building on adaptive random testing~\cite{chen2004adaptive}, which samples candidates that maximize their minimum distance to executed seeds to preserve diversity. 
These candidates are then refined using Stein Variational Gradient Descent (SVGD)~\cite{liu2016stein}, which we adapt and re-implement within our framework to approximate a posterior over the initial states from ARSG: the prior encodes plausible initial conditions, the likelihood quantifies the propensity of a condition to induce a failure, and the posterior concentrates on high-risk regions while maintaining coverage.
Each \textbf{particle} is a parameter vector that describes a dynamic object's initial state; SVGD applies gradient attraction toward high-risk regions and kernel repulsion to spread out (prevent collapse), uncovering multiple disjoint failure clusters in the high-dimensional search spaces and addressing GA's premature convergence. Refined seeds are fed back to the seed pool and simultaneously passed to the online tester to instantiate scenarios and trigger safety violations.

The online tester can be any existing online testing method, such as RL-based methods~\cite{liang2023rlaga} and gradient-based methods \cite{hanselmann2022king}.
This is because \tool is designed as a \textbf{plug-and-play} ADS testing framework that treats existing online testing methods as plug-ins.
In \tool, the \textit{offline seeder} is responsible for generating high-quality and diverse 
initial seeds that define the starting states of dynamic objects. 
The \textit{online tester} then
utilizes these seeds to produce concrete trajectories through simulation, exposing ADS safety issues under various and realistic conditions. 
They interface via gradient signals supplied by a learning-based hazard model that is trained using the online interaction data collected from the online tester to estimate failure likelihood from initial conditions.
Importantly, this process is iterative: feedback from 
the online stage guides subsequent seeding, progressively improving both the coverage of failure 
scenarios and the fidelity of the testing pipeline.
%\tool's offline seeder acts as the vehicle's engine---producing high-quality, diverse initial seeds---whereas the online tester serves as the wheels that translate seeds into concrete trajectories and outcomes. 
% This modular design allows the seeder to be placed upstream of any tester to enhance failure-finding effectiveness and broaden the diversity of uncovered failure modes, all without modifying the tester. 
% The whole system iterates until a stopping criterion is met (e.g., time budget or number of failures found).

In summary, our main contributions are as follows:

\begin{itemize}[noitemsep,topsep=0pt]
    \item 
    We introduce \tool, a plug-and-play ADS testing framework for dense, heterogeneous traffic that augments existing online testers to improve failure-finding effectiveness and diversity.
    \item We propose an SVGD-based seeding method that treats seeding as posterior inference—using a prior over plausible states, a learned hazard likelihood trained with online testing, and SVGD updates (gradient attraction + kernel repulsion) to target multiple high-risk regions while preserving diversity.
    \item We empirically evaluate \tool in CARLA across multiple ADS and testers, showing consistent gains in failure-finding effectiveness and coverage: it uncovers substantially more safety violations and a wider variety of failure modes, improving baseline testers rather than replacing them.
\end{itemize}

\section{Related Work}
\label{RW}
% \subsection{Multi-Agent Reinforcement Learning}

\subsection{ADS Testing via Offline Seed Generation}
\par Substantial effort has been devoted to testing ADS using offline seed generation followed by online scenario execution in simulation. 
In this setting, initial test configurations are generated offline, and the scenarios are executed in a simulator where dynamic objects evolve during runtime, but their trajectories remain pre-defined, limiting realism and adaptability.

\textit{Data-driven} approaches~\cite{rowe2025scenario, sun2023drivescenegen, gambi2019generating, hauer2020clustering, de2017assessment} leverage deep learning pipelines to synthesize realistic initial scenes and agent behaviors from real-world datasets. These methods excel at producing diverse, plausible simulation environments for training and evaluation; however, they largely emphasize open-ended scene generation and do not explicitly target the systematic triggering of safety violations in deployed ADSs. 

\par \textit{Combinatorial} methods~\cite{li2024generalization, kluck2018using, birkemeyer2022feature} systematically enumerate combinations of input parameters or features to ensure broad coverage, but rely exclusively on offline techniques—such as ontology-based modeling and pairwise/interaction-based sampling. 

\par \textit{Search-based} approaches~\cite{li2020av, panichella2015reformulating, abdessalem2018testing, ben2016testing, ebadi2021efficient, huai2023sceno, luo2021targeting, cheng2024decictor, chen2024misconfiguration, huai2023doppelganger, liang2023rlaga, huai2023doppelganger} use evolutionary algorithms (e.g., GA) to randomly generate chromosome representations, execute them in a simulator, and employ multi-objective fitness functions to select high-fitness parent scenarios; crossover and mutation then yield offspring with potentially higher risk. While effective at identifying safety violations, these methods often exhibit limited diversity in the violations they uncover due to the inherent convergence behavior of evolutionary search.

\par \tool\ is a search-based testing framework that combines offline seed generation with online scenario exploration. 
Unlike prior approaches that rely solely on pre-defined or GA-based seeding, \tool\ integrates adaptive random seed generation with an SVGD-guided refinement, enabling both diversity and targeted exploration of high-risk regions. 
This design allows \tool\ to expose a broader range of safety violations while ensuring coverage across heterogeneous traffic scenarios.

\subsection{ADS Online Testing}
We use the term \textit{online testing} to refer to methods where dynamic objects’ trajectories are generated and adapted on the fly during simulation through interactions with the ego vehicle and environment. 
This differs from testing using only offline seed generation, where initial conditions are fixed in advance and trajectories remain pre-defined. 
Online testing approaches~\cite{lu2022rgchaser,koren2018adaptive,lu2022learning,feng2023dense,liang2023rlaga,hanselmann2022king} employ reinforcement learning or gradient-based optimization to manipulate dynamic objects or environmental factors (e.g., weather) in real time within an episode. 

Online testing methods are effective, but without failure-inducing initial scenarios the search space remains enormous. 
\tool\ addresses this challenge by combining adaptive random testing (ART) for diverse seed generation with an SVGD-guided refinement that concentrates seeds in high-risk regions. 
The resulting scenarios are then passed to the online testing component for interaction-based exploration, enabling the discovery of more numerous and diverse safety violations in ADS.

\section{\tool Approach}
\label{method}

This section presents the design of \tool. We first provide an overview of the framework, followed by detailed descriptions of its offline seeding and online testing.

\begin{figure*}[t]
    \centering
    \includegraphics[width=1\linewidth]{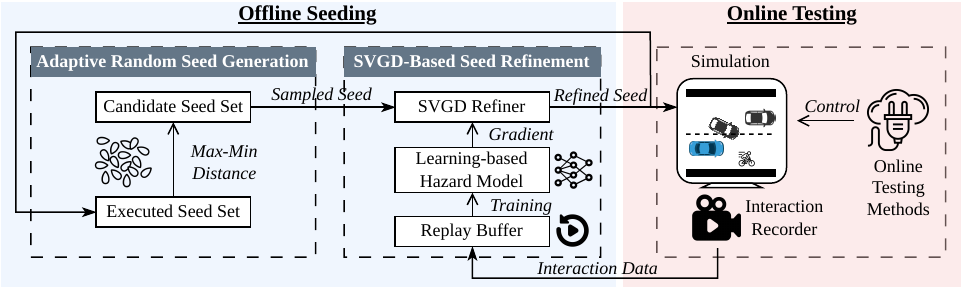}
    \vspace{-5mm}
	\caption{Framework overview of \tool.}
    \vspace{-5mm}
\label{overview}  
\end{figure*}

\subsection{Overview}

Figure~\ref{overview} presents the overall framework of \tool, which comprises an offline seeder and an online tester that operate iteratively to improve one another. The offline seeder consists of two stages, Adaptive Random Seed Generation (ARSG, \S\ref{Generation}) and SVGD-based Seed Refinement (\S\ref{Refinement}), that jointly produce diverse, failure-inducing test seeds. The online tester (\S\ref{online}) executes these seeds in a simulator using established online testing algorithms (e.g., reinforcement learning and gradient-based search) and returns hazardous-scenario feedback to the offline seeder.

The process begins with ARSG, which selects a seed from a candidate pool to maximize diversity relative to previously executed seeds. The selected seed is then refined via SVGD, which updates the associated particle ensemble to generate initial conditions for dynamic objects that are both more diverse and more likely to induce failures. Next, the refined seed is passed to the online tester, which manipulates the dynamic objects to elicit safety violations; simultaneously, the seed is also added to the set of executed seeds maintained by ARSG. During each episode, an interaction recorder logs interactions (e.g., speed, positions and headings) between each dynamic object and the ego vehicle and appends them to a replay buffer. Upon completion of the episode, the hazard model is updated using the replay buffer and its gradients are supplied to the SVGD refiner to guide the refinement of the next seed.

\subsection{Adaptive Random Seed Generation}
\label{Generation}

% \begin{figure}[h!]
%     \centering
%     \includegraphics[width=1\linewidth]{img/chromosome.jpg}
% 	\caption{Chromosome representation of the driving scenario}
% \label{Chromosome}  
% \end{figure}

\subsubsection{Seed Modeling}
\label{model_test_scenario}

To systematically encode the initial conditions of a test scenario, we adopt a chromosome-based representation. This representation comprises two primary components: the ego vehicle's route gene and the dynamics gene. The ego vehicle's route gene encodes the ego vehicle's initial orientation and position, as well as its destination, which is defined as the farthest valid waypoint within 200 meters along the vehicle's initial heading. This approach ensures consistency in test case length and comparability across scenarios. The dynamics gene specifies the type of each dynamic object present in the scenario, together with their respective initial positions and orientations. This structured encoding facilitates systematic scenario generation and supports comprehensive exploration of the scenario space.

\subsubsection{Adaptive Random Seed Generator}
To facilitate comprehensive exploration of the scenario space, we propose the Adaptive Random Seed Generator (ARSG), which is designed to generate highly diverse test seeds. The ARSG adopts the Adaptive Random Testing (ART) strategy~\cite{chen2004adaptive}, which systematically enhances diversity among generated seeds. Specifically, ARSG maintains a set of candidate seeds and, at each iteration, selects the candidate that \textbf{maximizes the minimum distance to all previously executed seeds}. Once a seed is selected and executed in the simulator, it is added to the set of executed seeds.

Formally, let the set of executed seeds be
\[
T = \{t_1, t_2, \dots, t_n\}
\]
and the set of candidate seeds be
\[
C = \{c_1, c_2, \dots, c_k\}.
\]
The ARSG selects the next seed \(c^*\) according to
\[
c^* = \arg\max_{c_j \in C}\left( \min_{t_i \in T} \text{dist}(c_j, t_i) \right),
\]
where \(\text{dist}(\cdot,\cdot)\) denotes a distance metric, typically Euclidean distance. The candidate seeds in \(C\) are generated randomly.

The motivation for this approach stems from the observation that purely random seed generation disregards the spatial distribution of seeds within the input domain, which can result in suboptimal failure detection—especially for failures that manifest in non-point patterns~\cite{chen2004adaptive}. By contrast, the ARSG explicitly promotes uniform spatial coverage, thereby increasing the probability of intersecting failure-prone regions that may be clustered or structured (e.g., as strips or blocks) within the input space. This uniformity is particularly advantageous for the subsequent seed refinement stage, as it provides a diverse and well-dispersed set of initializations, helping to avoid premature convergence to local optima and enabling more effective exploration of high-risk regions.

\subsection{SVGD-Based Seed Refinement}
\label{Refinement}
To effectively generate failure-inducing scenarios while maintaining diversity, we introduce a seed refinement approach based on Stein Variational Gradient Descent (SVGD). This approach integrates an interaction recorder and a replay buffer (\S\ref{Perception}) to systematically capture and store online interaction data from the online tester (\S\ref{online}). Leveraging this data, we train a learning-based hazard model (\S\ref{hazard_function}) that estimates the risk associated with different initial scenario configurations. The gradients provided by the trained hazard model are then utilized to guide the SVGD-based seed refiner (\S\ref{SVGD_INF}) toward regions of the scenario space with elevated risk, thereby enhancing the probability of uncovering hazardous cases. Simultaneously, the kernel-based repulsion mechanism inherent to SVGD fosters exploration across multiple modes and diverse regions of the scenario space, ensuring comprehensive and balanced coverage.

\subsubsection{Interaction Recorder and Replay buffer}
\label{Perception}

The interaction recorder and replay buffer are designed to capture and store online interaction data between the ego vehicle and surrounding dynamic objects during online testing. At each simulation frame \(t\), the system records the speed (\(v^{\text{ego}}_t\) for the ego vehicle and \(v^{(i)}_t\) for each dynamic object \(i\)), positions (\(p^{\text{ego}}_t\) and \(p^{(i)}_t\)), and headings (\(\psi^{\text{ego}}_t\) and \(\psi^{(i)}_t\)) of all relevant agents. This set of state information is continuously appended to the replay buffer, 
thereby facilitating subsequent analysis and enabling the learning of hazard models from the recorded interactions.

\subsubsection{Learning-based Hazard Model}
\label{hazard_function}

To estimate the probability that a specific initial configuration of the ego vehicle and surrounding dynamic objects will result in a safety violation within a finite time horizon, we employ a \emph{learning-based hazard model}. In the following, we first detail the construction of the feature vector that encodes the relative initial state between the ego vehicle and each dynamic object, and subsequently describe the derivation of learning targets from closed-loop simulation rollouts.

Formally, let $z \in \mathbb{R}^m$ denote a feature vector encoding the relative initial state between the ego vehicle and a dynamic object. For each dynamic object $i$ at the initial frame ($t{=}0$), we construct a five-dimensional, ego-centric feature vector $z_i \in \mathbb{R}^5$ capturing relative geometry and lane context:
\[
z_i=\big[
\underbrace{\Delta s_i}_{\text{longitudinal}},\;
\underbrace{\Delta d_i}_{\text{lateral}},\;
\underbrace{\cos\Delta\psi_i}_{\text{heading 1}},\;
\underbrace{\sin\Delta\psi_i}_{\text{heading 2}},\;
\underbrace{\lambda_i}_{\text{lane overlap}}
\big].
\]
Here, $(\Delta s_i, \Delta d_i)$ represent the longitudinal and lateral offsets of dynamic object $i$ in the ego frame at $t{=}0$, and $\Delta\psi_i = \psi^{(i)} - \psi^{\text{ego}}$ denotes the relative yaw. The variable $\lambda_i \in [0,1]$ quantifies the degree to which dynamic object $i$ overlaps with the ego’s current lane, where $\lambda_i = 1$ indicates perfect alignment within the same lane, and $\lambda_i = 0$ indicates complete absence from the ego’s lane. All trigonometric features are expressed in radians; the use of both $\cos\Delta\psi_i$ and $\sin\Delta\psi_i$ ensures continuity and avoids angle wrap-around issues. Each normalized component is clipped to $[-1,1]$ to enhance training stability. Notably, we omit kinematic features, as all dynamic objects are initially static.

\paragraph{Learning targets from closed-loop rollouts.}
Supervision based solely on collision events is excessively sparse for the driving domain: while most episodes do not result in collisions, many contain near-miss interactions that are highly informative for search. To address this, we design a dense, hand-crafted hazard target for each dynamic object, which serves as ground truth for online training of the lightweight hazard model. This hazard model provides smooth and informative gradients to guide SVGD. The design of the learning target adheres to three principles: (i) capturing both the relative position of the ego vehicle with respect to a dynamic object and the rate at which the ego is closing in; (ii) preserving a strict notion of failure for ego-responsible collisions; and (iii) aggregating risk over a short, critical time window to mitigate noise.

Given an episode of length $H$ with frames $t = 0, \ldots, H$, we extract the states of the ego and dynamic objects to construct an ego-centric time series.
Let
$d_t^{(i)}=\lVert p^{\text{ego}}_t-p^{(i)}_t\rVert_2$ be the Euclidean distance between the ego and dynamic object $i$. We define the \emph{unit bearing} (line of sight direction)
$\hat r_t^{(i)}=\frac{p^{(i)}_t-p^{\text{ego}}_t}{\lVert p^{(i)}_t-p^{\text{ego}}_t\rVert_2}$ and the instantaneous closing speed
\[
v^{\text{close}}_t(i)= -\,\hat r_t^{(i)\top}\!\big(v^{(i)}_t - v^{\text{ego}}_t\big),
\quad\text{so } v^{\text{close}}_t(i)\ge 0 \text{ indicates approach.}
\]
Using the \emph{unit} bearing isolates direction from range, making $v^{\text{close}}_t(i)$ dimensionally correct (m/s) and interpretable as the time derivative of distance along the line of sight.

We locate the most critical moment for dynamic object $i$ by
$t_i^\star=\arg\min_t d_t^{(i)}$ and consider a short 5-frames window 
$\mathcal{W}_i=\{t_i^\star-w,\ldots,t_i^\star+w\}$.
Within $\mathcal{W}_i$ we compute bounded, $[0,1]$-normalized indicators:
\[
s^{\text{dist}}_{t,i}=e^{-\bar d_i},\qquad
s^{\text{head}}_{t,i}=\tfrac{1-\cos\Delta\psi^{(i)}_t}{2},\qquad
s^{\text{close}}_{t,i}=\operatorname{clip}\!\left(\frac{\bar v^{\text{close}}_i}{v_{\max}},\,0,\,1\right).
\]
Here, $\Delta\psi^{(i)}_t$ is the relative yaw between the ego and dynamic object $i$ at time $t$, $\bar d_i$ is the average distance over $\mathcal{W}_i$, $\bar v^{\text{close}}_i$ is the average closing speed over $\mathcal{W}_i$, and $v_{\max}$ is the maximum velocity observed in the scenario. Intuitively, $s^{\text{dist}}$ is the normalized distance between the ego vehicle and the dynamic object and increases as the pair draws closer; $s^{\text{head}}$ is the normalized relative heading between the ego vehicle and the dynamic object and increases when their headings are aligned; and $s^{\text{close}}$ is the clipped closing speed of the ego vehicle relative to the dynamic object and increases when the ego is moving toward the dynamic object.

Each indicator is treated as an independent “no-risk” factor, and their contributions are aggregated multiplicatively across the window. For numerical stability and compactness, we perform this aggregation in log-space:
\[
S_i=
\exp\!\Biggl(
  \sum_{t\in\mathcal{W}_i}\!\Bigl[
    \log\!\bigl(1-\tilde s^{\text{dist}}_{t,i}\bigr)
   +\log\!\bigl(1-\tilde s^{\text{head}}_{t,i}\bigr)
   +\log\!\bigl(1-\tilde s^{\text{close}}_{t,i}\bigr)
  \Bigr]
\Biggr),
\]
where $\tilde s$ denotes the normalized cue values.

The resulting $S_i$ approximates the probability of \emph{no hazard} for dynamic object $i$ over the window $\mathcal{W}_i$; consequently, $y_i = 1 - S_i$ serves as a smooth near-miss score for the episode. In the event of a collision, $y_i \approx 1$; for episodes with no significant interaction, $y_i \approx 0$. We retain all samples, including those with $y_i \approx 0$, to address class imbalance and improve model calibration. Each feature vector $z_i$ is paired with its corresponding label $y_i$ to form the training data for the hazard model. This protocol yields dense, differentiable supervision for near-misses, enabling the surrogate model to provide informative gradients for SVGD-based search at the onset of each episode.

We train a hazard model $h_{\theta} : \mathbb{R}^m \to [0,1]$ that maps an ego-centric feature vector $z$ to a calibrated hazard score. The model is updated online after each episode using a binary cross-entropy loss between $h_{\theta}(z)$ and $y$.

The motivation for this design is that \textbf{SVGD updates particles via gradients}; thus, a differentiable hazard model is essential. A purely hand-crafted, non-differentiable proxy cannot provide gradients, and recomputing heuristic labels would necessitate re-running closed-loop simulations for each proposal, which is computationally prohibitive. The learned surrogate enables efficient, one-shot, differentiable hazard estimation at the beginning of each episode, thereby facilitating effective gradient-based search.

Classical threat metrics like TTC and THW work only in simple car-following settings and become undefined or unstable in common ADS-testing situations involving pedestrians, cyclists, lateral merges, or zero/negative closing speeds. They therefore cannot serve as a universal, differentiable training signal for our gradient-based pipeline. Our near-miss label captures the same underlying factors—distance, orientation, and closing speed—while remaining smooth, well-defined, and applicable across all interaction types, enabling stable gradient flow to the hazard model and SVGD. Thus, it is more suitable than traditional threat metrics for heterogeneous ADS testing.

% \[
% \min_{\theta}\;\frac{1}{|\mathcal{D}|}\!\sum_{(z,y)\in\mathcal{D}}
% \underbrace{\mathrm{BCE}\!\big(h_{\theta}(z),\,y\big)}_{\text{hazard calibration}}
% \;+\;\lambda\lVert\theta\rVert_2^2
% \;+\;\alpha\,\mathcal{R}_{\text{smooth}},
% \]
% where $\mathcal{R}_{\text{smooth}}$ is a small Jacobian penalty (encouraging local Lipschitzness) to stabilize the gradients that will be consumed by SVGD. In practice we normalize inputs, use a compact MLP with ReLU activations, apply light gradient clipping, and optionally perform temperature scaling for probability calibration.

% \paragraph{Use in SVGD-based search.}
% During test generation, particles $\{z_i\}$ represent candidate initial conditions (e.g., $(\Delta s,\Delta d,\Delta\psi)$ and local context) for multiple \emph{dynamics}. We evaluate $h_{\theta}(z_i)$ as the score and obtain $\nabla_{z}h_{\theta}(z_i)$ by automatic differentiation. Stein Variational Gradient Descent (SVGD) then moves particles using an attractive term driven by these gradients and a repulsive kernel to preserve diversity. Because $h_{\theta}$ is differentiable, SVGD acquires informative, low-variance directions without brittle finite differences on heuristic objectives.

% \paragraph{Online updates and replay.}
% After each episode, the new $(z,y)$ pairs—including near-misses—are appended to a replay buffer and used for brief fine-tuning. We maintain an exponential moving average of weights for stability and refresh the surrogate periodically so that its gradients reflect the latest operating regime.

% \paragraph{Why a learning-based surrogate?}

\subsubsection{SVGD Refiner}
\label{SVGD_INF}
To generate diverse and failure-inducing offline seeds, we employ an SVGD refiner to refine the seeds produced by the Adaptive Random Seed Generator. SVGD is a deterministic, particle-based method that approximates a target distribution by iteratively transporting a finite set of particles. At each iteration, particles move along a learned velocity field derived from Stein’s identity and a positive-definite kernel. This yields two complementary effects: an \emph{attraction} that drives particles toward high-probability regions via the target’s score, and a \emph{repulsion} that spreads particles across multiple modes to preserve diversity.

In our setting, the failure-inducing initial \emph{relative geometry} between the ego vehicle and dynamic objects forms a distribution. Given the learned hazard model $h_\theta$, our goal is to efficiently move a small set of candidate relative initial conditions toward failure-prone regions while maintaining diversity. Under our settings, a dynamic object is represented as a particle in SVGD, encoded by the mutable subset of the ego–dynamic-object relative features $x$:
\[
x \;=\; (\Delta s,\Delta d,\Delta\psi)\in\Omega,\qquad
\Omega \;=\; \{\,|\Delta s|\!\le D_s,\; |\Delta d|\!\le D_d,\; |\Delta\psi|\!\le \Psi_{\max}\,\},
\]
where $\Delta s$ is the relative longitudinal distance between the ego vehicle and the dynamic object, $\Delta d$ is the relative lateral distance, and $\Delta\psi$ is the relative heading; $D_s$, $D_d$, and $\Psi_{\max}$ are map-specified constants, with $D_s$ denoting the longitudinal range of the map, $D_d$ the lateral range, and $\Psi_{\max}$ the maximum heading difference, which equals $\pi$. Following the SVGD view~\cite{liu2016stein}, we define a target density $\pi (x)$

\[
\pi(x) \;\propto\; \exp\!\big(\tau\,h_\theta(z)\big)\,\mathbb{1}[x\in\Omega],
\]
where $\tau\!>\!0$ is a temperature to control the strength of the gradient. Intuitively, $\nabla_x \log \pi(x) = \tau,\nabla_x h_\theta(z)$ \textbf{pulls particles toward high-hazard modes (efficiency)}, whereas the SVGD kernel term \textbf{enforces spread (diversity)}. 

Let ${x_i}_{i=1}^N$ be the particles (one per selected dynamic object). SVGD transports the empirical measure by the velocity field
\[
\phi(x_i) \;=\; \frac{1}{N} \sum_{j=1}^N \Big[\, k(x_j,x_i)\,\underbrace{\nabla_{x} \log \pi(x_j)}_{\tau\,g_j}
\;+\;\beta\,\nabla_{x_j} k(x_j,x_i) \Big],
\]
where $g_j=\nabla_{x} h_\theta(z(x_j))$ and $\beta\!\in\!(0,1]$ weights repulsion. We use an anisotropic RBF kernel~\cite{liu2016stein}
\[
k(x,x') \;=\; \exp\!\Big(-\tfrac{1}{h}\,\|x-x'\|_{\Lambda}^2\Big),
\]
 The bandwidth $h$ is set by the median heuristic on pairwise $\|x_i-x_j\|_{\Lambda}^2$.
Each iteration performs
\[
x_i \;\leftarrow\; \Pi_{\Omega}\!\Big(x_i \;+\; \varepsilon\,\phi(x_i)\Big),
\]
with stepsize $\varepsilon$ and projection $\Pi_\Omega$ clipping to the box $\Omega$. Together, these effects attract particles to high-hazard regions while maintaining diversity.

% \paragraph{Selection, constraints, and rule-based guard.}
At the start of each episode we score all dynamics using $h_\theta$ and select the top-$K$ as particles. After each SVGD step, we apply a \emph{hard minimum-separation} guard in $(\Delta s,\Delta d)$: if any pair $(i,j)$ violates
$\sqrt{(\Delta s_i-\Delta s_j)^2+(\Delta d_i-\Delta d_j)^2}\ge r_{\min}$,
we push them apart along their local line of centers and re-clip to $\Omega$, where $r_{\min}$ equal to the lane width.
This complements the kernel repulsion with a physically motivated constraint, preventing unrealistically clustered initializations.

\paragraph{Advantages of Our Offline Seeding Approach.} In contrast to single-trajectory gradient ascent, SVGD simultaneously evolves a \emph{set} of candidate seeds, leveraging a principled repulsion mechanism that effectively balances exploitation—by directing particles toward high-hazard regions as indicated by $h_\theta$—and exploration—by maintaining coverage across multiple distinct modes. The integration of SVGD with ARSG further amplifies these benefits: while ARSG broadens the search to encompass a wider range of failure modes, SVGD enables more thorough exploration within each identified mode. This synergy is particularly valuable under stringent testing budgets, as it accelerates convergence toward diverse, failure-inducing initializations compared to random search or heuristic-based methods. Moreover, the incorporation of guard and box constraints ensures that all generated seeds correspond to physically plausible scenario configurations.

\subsection{Online Testing}
\label{online}

In our framework, \textit{online testing} denotes the process of dynamically modifying the trajectories of dynamic objects \textit{during simulation}, on a step-by-step basis, rather than relying on static, predefined trajectories. This approach enables the testing process to adapt in real time to the behavior of the system under test, as inputs are generated interactively within the simulation loop. To facilitate this, \tool offers a plug-in API that supports seamless integration with a variety of existing online testing methodologies. 

As a representative instantiation, we implement a gradient-based online testing technique inspired by KING~\cite{hanselmann2022king}. KING leverages a differentiable bicycle-kinematics model to compute gradients that guide the optimization of dynamic objects' trajectories, where even minor adjustments to steering or acceleration can significantly influence the probability of a collision. In our approach, the hazard model serves as an effective source of gradient information for online trajectory optimization: by performing gradient ascent, control updates are iteratively applied to each dynamic object's position and orientation to maximize the predicted hazard value.

We formally define the control space as follows. For vehicles and bicycles, the control input is a two-dimensional tuple \((\text{throttle}, \text{steering})\), representing the incremental changes in throttle and steering angle, respectively. For pedestrians, the control is a two-dimensional tuple \((x, y)\), corresponding to the longitudinal and lateral displacements. Our implementation generally follows KING's design for online control; further details regarding the mapping from hazard values to final control actions are provided in the \href{https://github.com/lfeng0722/PtoP/blob/main/FSE_2026_Supplementary.pdf}{\textbf{Section 3 of the supplementary material}}.

\section{Experimental Evaluation}
\label{exp}

In this section, we aim to evaluate the effectiveness of \tool in efficiently generating diverse safety violation scenarios under dense and heterogeneous traffic conditions. Specifically, we benchmark \tool against three state-of-the-art ADS testing methods across multiple map configurations and two advanced autonomous driving systems. To further elucidate the contributions of individual components within \tool, we conduct comprehensive ablation studies. Additionally, we perform a user study to systematically assess the realism of the safety violations identified by our approach. 

\subsection{Research Questions}

Our evaluation aims to answer the following research questions:

\begin{itemize}
    \item [RQ1] \textbf{What is the generalizability of \tool?} We examine whether \tool can interoperate with various online testing methodologies and assess its ability to generalize across diverse scenarios and autonomous driving systems.
    
    \item [RQ2] \textbf{How effective is \tool in identifying ADS safety violations compared to state-of-the-art techniques?} We evaluate whether \tool can discover a greater number and a wider variety of safety violations than existing approaches, while also considering the efficiency in terms of time budget.
    
    \item [RQ3] \textbf{What is the impact of individual components on the overall performance of \tool?} We analyze the contributions of the Adaptive Random Seed Generation and Stein Variational Gradient Descent components to the effectiveness of \tool.
    
    \item [RQ4] \textbf{How realistic are the safety violations identified by \tool?} We assess the plausibility and realism of the safety violations detected by \tool through a user study involving domain experts.
\end{itemize}

\subsection{Experimental Setup}
\label{setup}
We conduct our experiments using the high-fidelity CARLA simulator~\cite{dosovitskiy2017carla}, which provides comprehensive support for a range of autonomous driving systems (ADSs) and enables the simulation of both urban and rural traffic environments~\cite{dosovitskiy2017carla,zhong2022neural}. While our approach does not impose a theoretical upper bound on the number of dynamic objects, practical constraints such as computational resources and map dimensions lead us to configure each scenario with a fixed set of $20$ dynamic objects, consisting of vehicles, bicycles, and pedestrians. Each experimental configuration is executed for $400$ runs, with the entire process repeated four times to ensure statistical robustness; we report the averaged results across these repetitions.

\begin{definition}[Safety Violation]
    \label{def:safety_violation}
To rigorously assess safety, we need to identify the safety-critical behavior conduct by the ego vehicle. Safety-critical behavior refers to system actions or states that can directly lead to a safety violation. We define \textit{safety violations} within the test scenarios according to the following criteria:

\begin{itemize}[noitemsep,topsep=0pt]
    \item \textbf{I (Collision)}: The ego vehicle is involved in a collision with another vehicle.
    \item \textbf{II (Lane Departure)}: The ego vehicle crosses a solid lane marking.
    \item \textbf{III (Red Light Violation)}: The ego vehicle proceeds through a red traffic light.
    \item \textbf{IV (Motionless)}: The ego vehicle remains stationary for more than 15 seconds, despite the existence of a feasible path along its route.
\end{itemize}
\end{definition}

\subsection{Experimental Design}

\subsubsection{Generalizability (RQ1)}
We conducted experiments in the CARLA simulator across multiple scenarios and ADSs, integrating two online components. We conducted experiments across various maps of CARLA, each specifically selected to represent distinct driving scenarios:

\begin{itemize}[noitemsep,topsep=0pt]
    \item \textbf{Town01 (compact urban / dense traffic rules)}: A compact urban environment characterized by single-lane streets, tight intersections, and active traffic lights. This map is well-suited for assessing low-speed maneuvers in dense traffic and interactions among heterogeneous dynamic objects.
    \item \textbf{Town04 (highway-dominant / high-speed dynamics)}: A complex highway scenario featuring multiple lanes, overpasses, and merging lanes, making it ideal for evaluating high-speed driving and lane-changing behaviors.
    \item \textbf{Town07 (rural / navigation complexity)}: A large rural environment with varied road layouts, including unmarked roads without lane markings, and diverse route options. It features a richer road topology and higher navigation complexity, making it suitable for evaluating general driving policies, long-horizon navigation robustness, and interactions among heterogeneous dynamic objects.
    \item \textbf{Town10 (metropolitan / complex intersections + roundabouts)}: A metropolitan setting with dual-lane road layouts, roundabouts, and intersections, providing a challenging environment for testing urban driving policies and interactions among heterogeneous dynamic objects. As with Town04, traffic lights are not implemented in the Apollo version.
\end{itemize}

We evaluated three advanced ADSs within CARLA:
\begin{itemize}[noitemsep,topsep=0pt]
    \item \textbf{Apollo 8.0} \cite{Apollo}: Apollo represents a widely used, production-oriented autonomy stack with a full pipeline (localization/perception/planning/control). Evaluating on Apollo increases realism relative to CARLA’s built-in controllers because failures may emerge from \emph{multi-module interactions} (e.g., planning–control coupling) that do not exist in rule-based baselines.
    \item \textbf{Autoware (0.49.0)} \cite{autowarefoundation_autoware}: Autoware is a major open-source ADS in the robotics/autonomous driving ecosystem. Including Autoware increases representativeness by covering a \emph{different architecture and planning/control design} from Apollo, and by reflecting a widely used platform for academic and industry prototyping.
    \item \textbf{Traffic Manager} \cite{dosovitskiy2017carla}: Traffic Manager provides a stable and reproducible baseline with deterministic behavior and minimal stack complexity. We include it to (i) separate improvements due to our method from failures attributable to complex perception/localization stacks, and (ii) quantify performance relative to a commonly used simulator-default ADS in prior CARLA-based testing work.
\end{itemize}

For the reinforcement learning (RL) component, we adapted the RL module from GARL~\cite{liang2023rlaga}, employing the DQN algorithm~\cite{mnih2013playing}. The state representation for each dynamic object consists of its relative position and orientation with respect to the ego vehicle, consistent with the $x$ defined in \\\ref{SVGD_INF}. To accommodate the heterogeneous nature of dynamic objects, we defined separate action spaces: one for vehicles and bicycles, and another for pedestrians. The action space for vehicles and bicycles includes seven discrete actions: \textbf{lane keeping}, \textbf{accelerate}, \textbf{brake}, \textbf{right-lane change with acceleration}, \textbf{right-lane change with deceleration}, \textbf{left-lane change with acceleration}, and \textbf{left-lane change with deceleration}. For pedestrians, the action space comprises four actions: \textbf{forward}, \textbf{backward}, \textbf{left}, and \textbf{right}. The reward function is defined as the negative distance between the dynamic object and the ego vehicle, thereby incentivizing the agent to approach the ego vehicle. Each RL agent was pre-trained for 1,000 episodes on each map.

The generalizability of \tool\ is quantitatively assessed using the \textbf{Safety Violation Rate}, which measures the proportion of test cases in which the ego vehicle experiences a safety violation (as we defined in Definition~\ref{def:safety_violation}) during testing.
Note that we do not include cross-map evaluation. This is because our approach does not involve any pre-training process; instead, all components are trained online during the testing phase.

%\jz{I suggest categorizing them as collision failure and timeout failure, as both pose safety risks. A timeout failure occurs when the ego vehicle fails to reach its destination within XX minutes, which is XX times the normal duration. This typically indicates that the ego vehicle has stopped moving, creating potential hazards for other vehicles on the road.}

\subsubsection{Effectiveness (RQ2)}

In RQ2, we evaluate \tool against three state-of-the-art (SOTA) baselines on the same CARLA maps used in RQ1, focusing on the most robust ADS identified in the RQ1 results. The selected SOTA baselines—\textit{MOSAT}~\cite{tian2022mosat}, \textit{GARL}~\cite{liang2023rlaga}, and \textit{KING}~\cite{hanselmann2022king}—cover both online and testing with offline seed generation techniques. 

\begin{itemize}[noitemsep,topsep=0pt]

    \item \textit{\textbf{MOSAT}}~\cite{tian2022mosat}: We implement the MOSAT baseline using the NSGA-II algorithm~\cite{deb2000fast}. The chromosome representation is consistent with that of \tool\ (see \S\ref{model_test_scenario}), except that the \textit{Dynamics Gene} encodes a discrete action sequence for each dynamic object. Each action sequence comprises 10 actions, selected from the same action space as the RL agent described in RQ1. Variation operators are applied as follows: upon mutation (according to a predefined rate), the chromosome representation is resampled; upon crossover, a single-point crossover is performed between the \textit{Dynamics Gene} segments of two chromosomes. The multi-objective fitness function optimizes for: (1) \textit{Safety Violation}, a binary indicator denoting whether a test case induces a safety violation; and (2) \textit{Diversity}, which quantifies the uniqueness of each chromosome as the average Euclidean distance to all other chromosomes in the generation. The diversity metric is computed as~\cite{lehman2011abandoning, lehman2008exploiting, liang2023rlaga}:
    \[
        D_i = \frac{1}{k} \sum_{j=1}^{k} d\left(x_i, x_j\right)
        \label{eq:parameter_distance-obj}
    \]
    where \( D_i \) denotes the diversity score for the \( i \)-th chromosome, \( x_i \) and \( x_j \) are chromosome representations, \( d(\cdot, \cdot) \) is the Euclidean distance, and \( k \) is the generation size. We reproduce the action patterns and trigger conditions as specified in MOSAT~\cite{tian2022mosat}.

    \item \textit{\textbf{GARL}}~\cite{liang2023rlaga}: We adapt GARL from its original UAV marker-based landing context. In this baseline, multiple single-agent RL models independently control the surrounding vehicles, without inter-agent communication, within scenarios generated by a genetic algorithm. For the GA component, we employ NSGA-II~\cite{deb2000fast}, utilizing the same chromosome representation, variation operators, and multi-objective fitness function as in the MOSAT baseline. The RL component is implemented identically to that described in RQ1.

    \item \textbf{\textit{KING}}~\cite{hanselmann2022king}: We implement the online testing approach as detailed in \ref{online}, with each scenario initialized randomly.

\end{itemize}

% To evaluate whether \tool\ can generate more and more-diverse safety violations, we compare \tool \ with three SOTA baselines:
%  \item \textit{MOSAT}: This baseline is adopted from MOSAT \cite{tian2022mosat}, utilizing a predefined action sequence offline. The sequences are randomly initialized and optimized via NSGA-II \cite{deb2000fast} through crossover and mutation. The GA fitness function is identical to the GA component in \tool, excluding the objective related to ART (trigger time).
%     Once the surrounding vehicles enter a threshold distance to the ego vehicle ($L_{width}$), a corresponding action pattern is triggered.

%     \item \textit{GARL} : We adapt GARL~\cite{liang2023rlaga} from the UAV marker-based landing scenario. This baseline enables multiple single-agent RL models to independently control surrounding vehicles without communication. The agent's model structure and training environment remain the same as in \tool. The agent's objective is to approach the ego vehicle while maintaining a lane-width distance to match encirclement in \tool, we ensure training convergence.  The number and initial positions of surrounding vehicles are sampled using the same GA as \tool, excluding the objective related to ART (trigger time).

We use following metrics to assess this RQ:
\begin{itemize}
     \item \textbf{Safety Violation Rate}: As defined in RQ1, this metric quantifies the proportion of test cases in which the ego vehicle experiences a safety violation.
    
    \item \textbf{TOP-10}: This metric measures the number of test rounds required to identify the first ten safety violations, thereby reflecting the efficiency of each method in uncovering critical failures.
    
    \item \textbf{Parameter Distance}: Inspired by novelty search~\cite{lehman2011abandoning,lehman2008exploiting}, this metric quantifies the diversity among the generated safety violations. A higher value indicates greater diversity. The metric is formally defined as:
    \[
        \rho(x) = \frac{1}{n} \sum_{i=1}^n \left[ \frac{1}{n} \sum_{j=1}^n d(x_i, x_j) \right]
        \label{eq:parameter_distance}
    \]
    where $x_i$ and $x_j$ represent the scenario representation vectors of the $i$-th and $j$-th safety-violating cases, respectively. Each vector encodes the positions and orientations of the ego vehicle and dynamic objects. The function $d(x_i, x_j)$ denotes the Euclidean distance, and $n$ is the total number of generated safety violations. All gene values are normalized to the range $[0,1]$. Notably, the maximum possible distance between two chromosome representations is $1$.
    
    \item \textbf{Map Coverage}: This metric assesses the fraction of map spawn points utilized by dynamic objects near the ego vehicle in safety-violating scenarios. For each safety violation, we collect the spawn positions of dynamic objects within a radius of \(R=50\,\mathrm{m}\) from the ego vehicle, map each position to its nearest spawn point within a threshold of $0.05\,\mathrm{m}$, and remove duplicates. Let \(\mathcal{S}_m\) denote the set of all map spawn points and \(\mathcal{U}_m \subseteq \mathcal{S}_m\) the subset matched by the above procedure. The map coverage is then computed as:
    \[
        \mathrm{Map Coverage} = \frac{|\mathcal{U}_m|}{|\mathcal{S}_m|} \times 100\%.
    \]
    
%     \item \tochecklf{\textbf{Behavior Diversity}: This metric measures how diverse the takeover behaviors of dynamic vehicles are across safety-violating test cases. For each safety violation test case, we record the per-tick action sequence for every \emph{takeover} vehicle (i.e., vehicles for which an explicit control action is issued). We then represent each test case by a fixed-length feature vector that aggregates all takeover action sequences in that case, including (i) action-token frequency counts and (ii) action-bigram (consecutive action-pair) frequency counts. After performing per-dimension min--max normalization to map each feature to $[0,1]$, the behavior diversity over all safety violation test cases is computed as:
% \[
%     D_b = \frac{1}{n} \sum_{p=1}^{n} \left[ \frac{1}{n} \sum_{q=1}^{n} d(\mathbf{v}_p, \mathbf{v}_q) \right],
%     \label{eq:behavior_diversity}
% \]
% where $n$ is the number of safety-violating test cases, $\mathbf{v}_p$ and $\mathbf{v}_q$ are the normalized action-sequence feature vectors of the $p$-th and $q$-th test cases, and
% \[
%     d(\mathbf{v}_p, \mathbf{v}_q) = \frac{1}{D}\sum_{k=1}^{D}\left|\mathbf{v}_p[k]-\mathbf{v}_q[k]\right|
% \]
% is the mean absolute difference over the $D$ feature dimensions. A larger $D_b$ indicates more diverse takeover behaviors across violating test cases.}

    \item \textbf{Trajectory Coverage}: To quantify the behavioral diversity of dynamic objects, we use \emph{trajectory coverage}. Since the online component involves both discrete and continuous action spaces, it is difficult to measure diversity directly in the action space (e.g., via distance-based metrics). Following \cite{liang2023rlaga}, we treat trajectories as the outcome of behaviors and compute coverage in a map-aware manner. Concretely, we collect all positions of each dynamic object during a run, map each position to its nearest waypoint within a threshold of $0.05\,\mathrm{m}$, and remove duplicates. Let \(\mathcal{S}_t\) denote the set of all map waypoints and \(\mathcal{U}_t \subseteq \mathcal{S}_t\) the subset matched by the above procedure. The trajectory coverage rate is computed as:
    \[
        \mathrm{Trajectory Coverage} = \frac{|\mathcal{U}_t|}{|\mathcal{S}_t|} \times 100\%.
    \]

    % \item \tochecklf{\textbf{Fault Taxonomy}: This metric provides a qualitative analysis of the causes of safety violations. We manually review all safety-violating test cases, identify their root causes, and summarize the findings into a fault taxonomy.}

    \item \textbf{Time Consumption}: This metric records the total time required by each method to exhaust its allotted test budget. While accuracy is paramount, excessive test case generation time can impede practical adoption in real-world testing pipelines.
    
    % Given road waypoints \( R = \{(x_i, y_i)\}_{i=1}^{M} \) and trajectory points \( T = \{(x_j, y_j)\}_{j=1}^{N} \),\jz{define M and N} a waypoint is considered covered if its nearest trajectory point, identified using a KD-Tree\jz{K-Dimensional tree? you need to give self-contained explanation}, is within a threshold \( \tau \).\jz{this part is confusing, KD-Tree is essnetialy a binary tree recursively partition the space and each node split plane along one dimension generating left subtree and right subtree, How does it work with this threshold? splitting threshold? If so, you mean if a waypoint is at the same subtree of its nearest trajectory point, so it is covered? be clear} The coverage ratio is defined as:
    % \begin{equation}
    % \text{Coverage Ratio} = \frac{|\{(x_i, y_i) \in R \mid \min\limits_{(x_j, y_j) \in T} d((x_i, y_i), (x_j, y_j)) < \tau\}|}{M},
    % \end{equation}
    % where \( d((x_i, y_i), (x_j, y_j)) \) is the Euclidean distance.

\end{itemize}

Additionally, we compare our method with TARGET~\cite{deng2023target}. Since TARGET follows a different testing methodology, a direct comparison under the same set of metrics is not applicable. Instead, we compare the two methods in terms of \textit{efficiency} in finding safety violations.

To ensure a fair comparison, we standardize the safety-violation definitions and focus on the overlap between the two methods: \textit{collision}, \textit{lane departure}, and \textit{motionless}. We perform this comparison on all maps used in RQ1 and report the overall result, with both Apollo and Autoware as the system under test. As an efficiency measure, we report the average number of violations of each type found within a fixed time budget of 6 hours (4 runs).

\subsubsection{Ablation Study (RQ3)}
To evaluate the contributions of individual components in \tool, we conduct an ablation study with two settings:

\begin{itemize}[noitemsep,topsep=0pt]
    \item \textbf{\textit{\tool\ with vs. without ARSG}}: ARSG is replaced with a random seed generator to assess its role in discovering additional failure modes exploitable by SVGD.
    
    \item \textbf{\textit{\tool\ with SVGD vs. with GA}}: SVGD and ART are replaced with a GA-based generator (same as MOSAT in RQ2) to compare ARSG+SVGD against GA in producing diverse, failure-inducing scenarios.
\end{itemize}

All experiments use the same maps, ADS configurations, and metrics as RQ2 for fair comparison.
%To rigorously evaluate the contributions of individual components within \tool, we conduct an ablation study under the following experimental settings:

%\begin{itemize}[noitemsep,topsep=0pt]
 %   \item \textbf{\textit{\tool\ with vs. without ARSG}}: In this configuration, the ARSG module is removed from \tool\ and replaced with a random seed generator. This comparison is designed to assess the extent to which ARSG facilitates the discovery of additional failure modes that can be further exploited by SVGD.
    
   % \item \textbf{\textit{\tool\ with SVGD vs. with GA}}: Here, both SVGD and ART are replaced with a genetic algorithm (GA)-based initial scenario generator, utilizing the same GA implementation as MOSAT described in RQ2. This setting enables a direct comparison of the effectiveness of ARSG+SVGD versus GA in producing initial scenarios that are both failure-inducing and diverse.
%\end{itemize}

%All experiments are conducted on the same set of maps and with the same ADS configurations as in RQ2. The evaluation employs identical metrics to ensure a fair and consistent comparison across all settings.

\subsubsection{Fidelity (RQ4)}

To validate the realism of the generated safety violations, we conducted a user study involving experienced drivers. We randomly selected 12 videos from our record and recruited 30 experienced drivers globally through Prolific~\cite{Prolific2024}. Participants, aged between 18 and 60 years and holding valid driver's licenses, represented diverse genders. Each participant evaluated the realism of surrounding vehicle trajectories by answering the question: \textit{``How realistic are the surrounding vehicles’ trajectories in this video?''} Responses were recorded using a Likert scale with the following categories: \textit{``Very unrealistic,'' ``Slightly unrealistic,'' ``Neutral,'' ``Slightly realistic,''} and \textit{``Very realistic.''} If a participant rated a video below \textit{``Neutral,''} they were asked to provide a reason for their rating, the survey can be found \href{https://docs.google.com/forms/d/e/1FAIpQLSdQWwm_bTLhRHKWUO2KnhxM6AtaqQxmf_L54-zmdtfCtCHwYw/viewform?usp=dialog}{\textbf{here}} (enter any worker ID). Additionally, we calculated Weighted Fleiss' Kappa \cite{deng2023target} to measure inter-rater agreement among participants. Detailed demographic information can be found \href{https://github.com/lfeng0722/PtoP/blob/main/FSE_2026_Supplementary.pdf}{\textbf{here}}. The user study interface and the Weighted Fleiss' Kappa calculation are provided in  \href{https://github.com/lfeng0722/PtoP/blob/main/FSE_2026_Supplementary.pdf}{\textbf{Sections 1 and 2 of the supplementary material}}.

\subsection{Generalizability of \tool \ (RQ1)}

\begin{table}[t]
    \centering
    \caption{Averaged ADS safety violation rate (\%) and across different CARLA maps when \tool \ using KING as online part.}
    % \vspace{-3mm}
    \scalebox{1}{
        \begin{tabular}{|c|c|c|c|c|}
            \hline
            \textbf{Map} & \textbf{Metric} & \textbf{Apollo} & \textbf{Autoware} & \textbf{Traffic Manager} \\
            \hline
            \multirow{1}{*}{\textbf{Town1}} 
                & Violation Rate (\%) & \textbf{23.06} & 28.68 & 28.64  \\
            \hline
            \multirow{1}{*}{\textbf{Town4}} 
                & Violation Rate (\%) & \textbf{18.31} & 22.17 & 21.47 \\
            \hline
            \multirow{1}{*}{\textbf{Town7}}
                & Violation Rate (\%) & \textbf{30.13} & 37.18 & 37.85 \\
            \hline
            \multirow{1}{*}{\textbf{Town10}} 
                & Violation Rate (\%) & \textbf{19.81} & 27.58 & 29.45 \\
            \hline
        \end{tabular}
    }
    \label{tab:general1}
\end{table}

\begin{table}[t]
    \centering
    \caption{Averaged ADS safety violation rate (\%) across different CARLA maps when \tool \ using RL as online part.}
    % \vspace{-3mm}
    \scalebox{1}{
        \begin{tabular}{|c|c|c|c|c|}
            \hline
            \textbf{Map} & \textbf{Metric} & \textbf{Apollo} & \textbf{Autoware} & \textbf{Traffic Manager} \\
            \hline
            \multirow{1}{*}{\textbf{Town1}} 
                & Violation Rate (\%) & \textbf{20.94} & 31.13 & 33.74  \\
            \hline
            \multirow{1}{*}{\textbf{Town4}} 
                & Violation Rate (\%) & \textbf{18.75} & 24.52 & 23.35 \\
            \hline
            \multirow{1}{*}{\textbf{Town7}}
                & Violation Rate (\%) & \textbf{28.00} &35.31 & 34.75 \\
            \hline
            \multirow{1}{*}{\textbf{Town10}} 
                & Violation Rate (\%) & \textbf{21.31} & 31.03 & 33.45 \\
            \hline
        \end{tabular}
    }
    \label{tab:general2}
\end{table}

Tables~\ref{tab:general1} and \ref{tab:general2} summarize the generalizability of \tool\ in CARLA across Town~1, Town~4, Town~7, and Town~10. We evaluate \tool\ with Apollo, Autoware, and Traffic Manager under different online methods.
Overall, both online methods integrate well with \tool\ and can uncover safety violations across all scenarios and ADSs. Online RL performs slightly better than KING in Town~4 and Town~10; however, KING outperforms RL in Town~1 and Town~7. This is likely because Town~1 and Town~7 are more rural, where most roads are single-lane; with RL’s discrete action space, lane-change actions are limited in these maps. In contrast, KING uses a gradient-based strategy to synthesize continuous actions, which gives it an advantage in Town~1 and Town~7.

% \textcolor{blue}{All methods—\tool and the SOTA baselines—generate 1,600 test episodes per map per agent (Apollo, Autoware, and Traffic Manager), ensuring a fair and identical evaluation budget. Among these, PtoP yields 369/293/317 violating cases in Town1/04/10 with Apollo; \tochecklf{please add the corresponding Autoware counts here}; in comparison, MOSAT yields 310/251/278, GARL yields 289/240/286, and KING yields 336/243/274, respectively. }

\tool\ consistently records fewer safety violations with Apollo, whereas Traffic Manager and Autoware exhibit higher and comparable violation rates. This is likely because Traffic Manager implements only basic autonomous-driving functions (e.g., stopping for vehicles ahead). In addition, Autoware may fail to detect static or slow-moving vehicles, which can cause the ego vehicle to collide with the vehicle in front; this behavior is consistent with a previously reported Autoware--CARLA issue~\cite{deng2023target}. Additionally, all ADS configurations show the fewest safety violations in Town~4, likely due to its significantly larger area, which reduces the frequency of interactions that lead to safety violations.

\par We conducted a case study to manually analyze the causes of selected safety-violation cases identified by \tool\ in Apollo. We selected cases that involve heterogeneous dynamic objects and exhibit diverse causes of safety violations. \textbf{The vehicle annotated with a red dot in the image is the ego vehicle.}

% 需要在导言区

\begin{figure}[!t]
  \centering

  %—— 左半行：两张图 + 小标题A ——%
  \begin{subfigure}[t]{0.49\linewidth}

    \centering
    \includegraphics[width=0.49\linewidth]{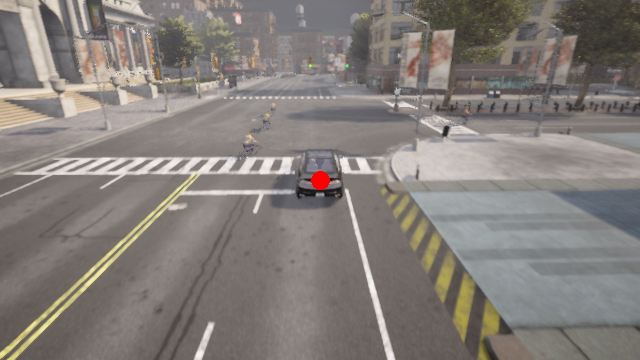}\hfill
    \includegraphics[width=0.49\linewidth]{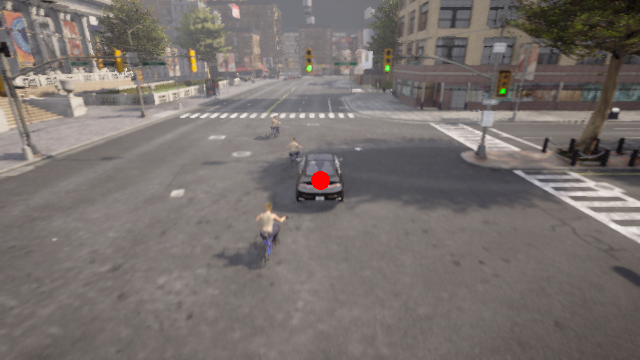}
  \subcaption{Case Study I}
  \end{subfigure}\hfill
  %—— 右半行：两张图 + 小标题B ——%
  \begin{subfigure}[t]{0.49\linewidth}
    \centering
    \includegraphics[width=0.49\linewidth]{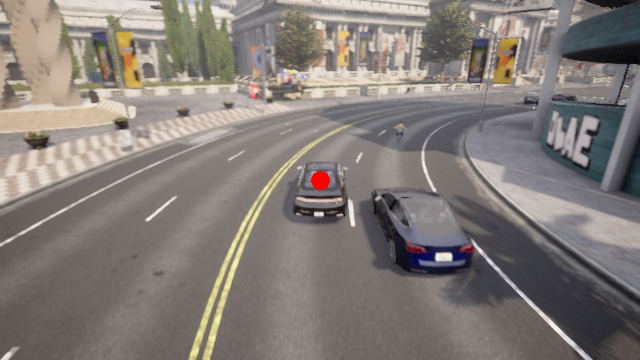}\hfill
    \includegraphics[width=0.49\linewidth]{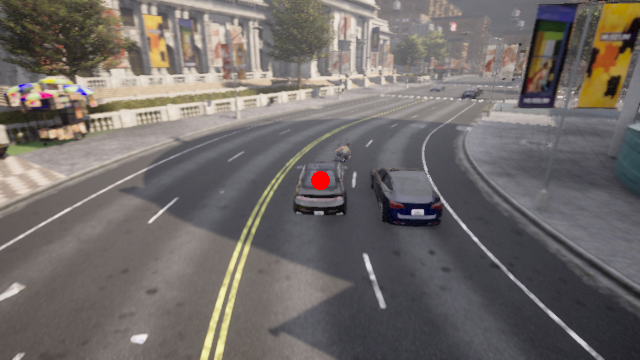}
    \subcaption{Case Study II}
  \end{subfigure}
     %—— 左半行：两张图 + 小标题A ——%
  \begin{subfigure}[t]{0.49\linewidth}
    \centering
    \includegraphics[width=0.49\linewidth]{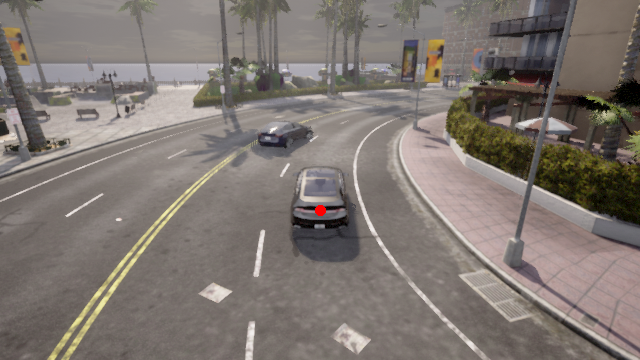}\hfill
    \includegraphics[width=0.49\linewidth]{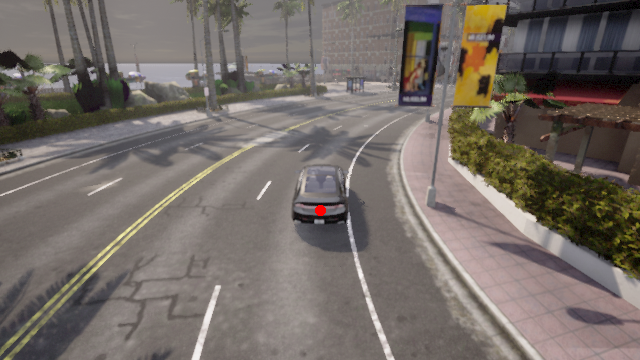}
    \subcaption{ Case Study III}
  \end{subfigure}\hfill
  %—— 右半行：两张图 + 小标题B ——%
  \begin{subfigure}[t]{0.49\linewidth}
    \centering
    \includegraphics[width=0.49\linewidth]{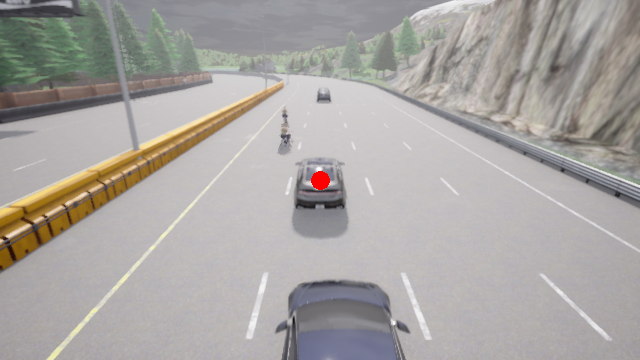}\hfill
    \includegraphics[width=0.49\linewidth]{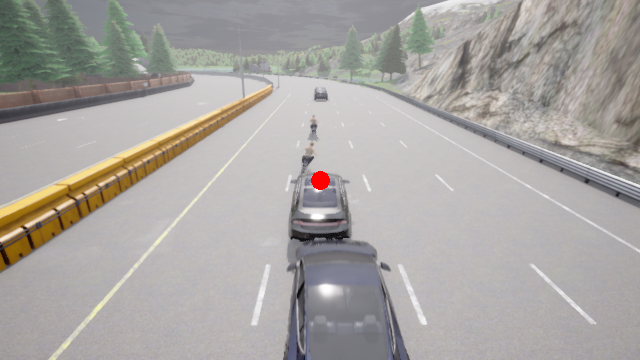}
    \subcaption{Case Study IV}
  \end{subfigure}
  \begin{subfigure}[t]{0.49\linewidth}
    \centering
    \includegraphics[width=0.49\linewidth]{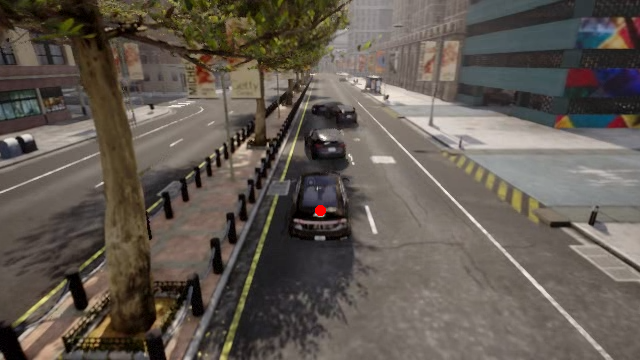}\hfill
    \includegraphics[width=0.49\linewidth]{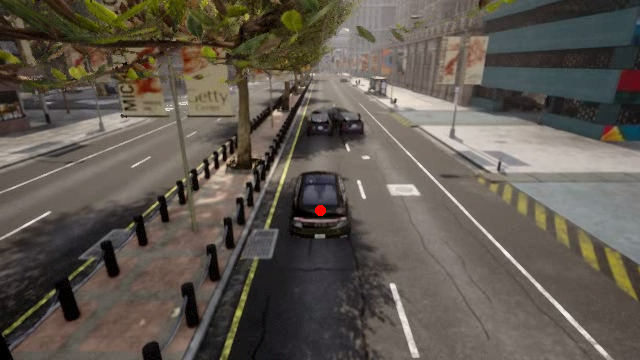}
    \subcaption{Case Study V}
  \end{subfigure}
   \begin{subfigure}[t]{0.49\linewidth}
    \centering
    \includegraphics[width=0.49\linewidth]{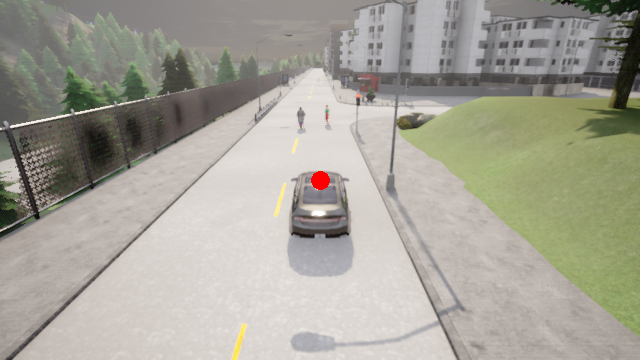}\hfill
    \includegraphics[width=0.49\linewidth]{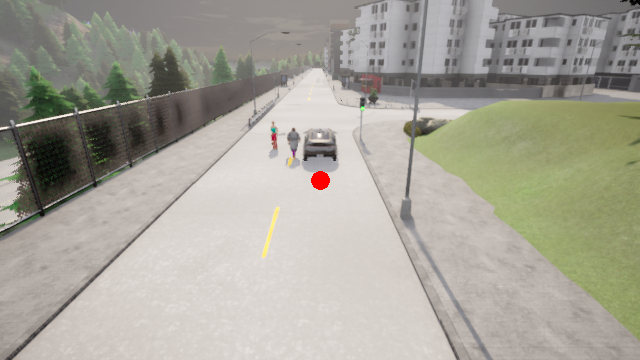}
    \subcaption{Case Study VI}
  \end{subfigure}

%   \begin{subfigure}[t]{0.49\linewidth}
%     \centering
%     \includegraphics[width=0.49\linewidth]{img/TOWN7/frame_315_marked.png}\hfill
%     \includegraphics[width=0.49\linewidth]{img/TOWN7/frame_316_marked.png}
%     \subcaption{\tochecklf{Case Study VII}}
%   \end{subfigure}
%   \vspace{-3mm}
  \caption{Case study.}
  \label{exp1}
\end{figure}

\subsubsection{Case Study I}
The ego vehicle collides with multiple cyclists due to unexpected lane changes by surrounding cyclists. In Figure \ref{exp1} (a), the ego vehicle and several cyclists are passing through an intersection when one cyclist accelerates to merge into the ego vehicle's lane, triggering a collision.

\subsubsection{Case Study II}

The ego vehicle collides with the cyclist due to acceleration of a following vehicle. As shown in Figure~\ref{exp1} (b), on a two-lane road, the ego vehicle is traveling in the left lane, a cyclist is ahead in the right lane, and another vehicle is driving behind the cyclist. When the following vehicle accelerates, the cyclist gives way by merging into the left lane (the ego vehicle’s lane). Due to the short headway, the ego vehicle collides with the cyclist.

\subsubsection{Case Study III}

% \begin{figure}[!t]
%     \centering
%     \begin{minipage}{0.49\linewidth}
%         \centering
%         \includegraphics[width=\linewidth]{img/TYPE4/IV1.png}
%     \end{minipage}
%     \hfill
%     \begin{minipage}{0.49\linewidth}
%         \centering
%         \includegraphics[width=\linewidth]{img/TYPE4/IV2.png}
%     \end{minipage}

%     \caption{Case Study III: Ego vehicle collides with the rear of multiple preceding surrounding vehicles.}
%     \label{exp3}
% \end{figure}

Figure~\ref{exp1} (c) illustrates a scenario where the ego vehicle crosses a solid line on the road. This occurs when a surrounding vehicle attempts to merge into the ego vehicle's lane while accelerating. Due to the close proximity between the ego vehicle and the surrounding vehicle, the ego vehicle attempts to avoid a collision by steering slightly to the right, inadvertently crossing the solid line.

\subsubsection{Case Study IV:}

% \begin{figure}[!t]
%     \centering
%     \begin{minipage}{0.49\linewidth}
%         \centering
%         \includegraphics[width=\linewidth]{img/TYPE3/III1.png}
%     \end{minipage}
%     \hfill
%     \begin{minipage}{0.49\linewidth}
%         \centering
%         \includegraphics[width=\linewidth]{img/TYPE3/III2.png}
%     \end{minipage}

%     \caption{Case Study IV: Ego vehicle crossed solid line on the road.}
%     \label{exp4}
% \end{figure}
Figure~\ref{exp1}(d) illustrates a scenario where the ego vehicle hits a cyclist. On a four-lane road, the ego vehicle is traveling with another vehicle following behind. A cyclist suddenly merges into the ego vehicle’s lane; the ego vehicle brakes sharply but still collides with the cyclist. Immediately after, the following vehicle rear-ends the ego vehicle.

% This occurs when a surrounding vehicle attempts to merge into the ego vehicle's lane while accelerating. Due to the close proximity between the ego vehicle and the surrounding vehicle, the ego vehicle attempts to avoid a collision by steering slightly to the right, inadvertently crossing the solid line.

% \subsubsection{Case Study V:}

% \begin{figure}[!t]
%     \centering
%     \begin{minipage}{0.49\linewidth}
%         \centering
%         \includegraphics[width=\linewidth]{img/TYPE5/V1.png}
%     \end{minipage}
%     \hfill
%     \begin{minipage}{0.49\linewidth}
%         \centering
%         \includegraphics[width=\linewidth]{img/TYPE5/V2.png}
%     \end{minipage}

%     \caption{Case Study V: Surrounding vehicle collides with the rear of ego vehicle.}
%     \label{exp5}
% \end{figure}

% Figure \ref{exp5} illustrates a scenario where a surrounding vehicle collides with the rear of the ego vehicle. The ego vehicle was waiting at a red light and began moving when the light turned green; however, its acceleration was slow. Meanwhile, a surrounding vehicle traveling on the same route at a higher speed approached from behind. After making a left turn, the surrounding vehicle was unable to decelerate in time and collided with the ego vehicle.

\subsubsection{Case Study V:}
% \begin{figure}[!t]
%     \centering
%     \begin{minipage}{0.49\linewidth}
%         \centering
%         \includegraphics[width=\linewidth]{img/TYPE6/VI1.png}
%     \end{minipage}
%     \hfill
%     \begin{minipage}{0.49\linewidth}
%         \centering
%         \includegraphics[width=\linewidth]{img/TYPE6/VI2.png}
%     \end{minipage}

%     \caption{Case Study V: Ego vehicle motionless.}
%     \label{exp6}
% \end{figure}

Figure \ref{exp1} (e) illustrates a scenario where the ego vehicle experiences a motionless state despite having a feasible route. Initially, two surrounding vehicles collide while attempting a lane change and come to a stop on the road. Later, a third surrounding vehicle attempts to overtake the ego vehicle by changing lanes with acceleration to find a new route. However, due to the obstructed view caused by the ego vehicle and the stationary collided vehicles ahead, the third vehicle collides with the already-stopped vehicles after completing its lane change. At this point, the ego vehicle has the option to take the right lane to continue its route but remains stuck on the road, ultimately leading to a motionless.

\subsubsection{Case Study VI:}

Figure~\ref{exp1} (f) illustrates a near miss with pedestrians. At the beginning, the ego vehicle is waiting at a red light while two pedestrians are crossing the road. When the light turns green, the ego vehicle starts moving; however, the pedestrian attempts to cross the road diagonally, and the ego vehicle does not slow down, accelerating toward the pedestrian—creating a highly dangerous situation.

\subsection{Effectiveness (RQ2)}

% ====================== Baseline ======================
\begin{table}[t]
\centering
\caption{Baseline results across towns. Best results are in bold.}
\vspace{-3mm}
\resizebox{\columnwidth}{!}{
\begin{tabular}{l l c c c c c c}
\hline
\textbf{Method} & \textbf{Town} & \textbf{Safety Violation \%} & \textbf{Top-10} & \textbf{Parameter distance} & \textbf{Map Coverage \%} & \textbf{Trajectory Coverage \%} & \textbf{Time Consumption (hours)} \\
\hline
\multirow{4}{*}{\textit{\tool}} 
 & Town1 & \textbf{23.06}\% & \textbf{39.5} & \textbf{0.331} & \textbf{31.17}\% & 51.4\% & 6 \\
 & Town4 & \textbf{18.31}\% & \textbf{62.5} & \textbf{0.317} & \textbf{20.53}\% & 61.4\% & 6 \\
 & Town7 & \textbf{30.13}\% & \textbf{42} & \textbf{0.319} & \textbf{79.65}\% & 90.3\% & 6 \\
 & Town10 & \textbf{19.81}\% & \textbf{45.25} & \textbf{0.315} & \textbf{63.54}\% & 89.2\% & 6 \\
\hline
\multirow{4}{*}{\textit{MOSAT \cite{tian2022mosat}}} 
 & Town1 & 19.38\% & 49.5 & 0.302 & 25.96\% & 38.6\% & 6 \\
 & Town4 & 15.69\%  & 70.5 & 0.292 & 15.31\%  & 40.6\% & 6 \\
 & Town7 & 24.00\% & 60.5 & 0.281 & 55.70\% & 69.2\% & 6 \\
 & Town10 & 17.38\% & 52.75 & 0.291 & 57.04\% & 70.3\% & 6 \\
\hline
\multirow{4}{*}{\textit{GARL \cite{liang2023rlaga}}} 
 & Town1 & 18.06\% & 52 & 0.303 & 25.22\% & 48.6\% & 6 \\
 & Town4 & 15.00\%  & 73 & 0.292 & 15.78\%  & 55.3\% & 6 \\
 & Town7 & 24.63\% & 71.25 & 0.287 & 56.48\% & 80.6\% & 6 \\
 & Town10 & 17.89\% & 58.75 & 0.288 & 54.41\% & 85.6\% & 6 \\
\hline
\multirow{4}{*}{\textit{KING \cite{hanselmann2022king}}} 
 & Town1 & 21.00\% & 45 & 0.302 & 25.25\% & 53.7\% & 6 \\
 & Town4 & 15.19\%  & 73 & 0.292 & 15.23\%  & 62.4\% & 6 \\
 & Town7 & 26.13\% & 53.25 & 0.291 & 75.18\% & 91.5\% & 6 \\
 & Town10 & 17.13\% & 56.50 & 0.292 & 56.31\% & 90.6\% & 6 \\
\hline

\end{tabular}}
\label{tab:exp2_baseline}
\end{table}

\begin{table}[t]
\centering

\label{tab:pvalues_towns}
\small
\caption{Two-sided paired \emph{t}-test results of baselines vs. \tool (n=4 runs): \emph{p}-values, 95\% confidence intervals (CIs), and effect sizes (Cohen’s $d$ for paired samples)}.
\vspace{-3mm}
\resizebox{\columnwidth}{!}{
\begin{tabular}{llccc ccc ccc ccc}
\toprule
\textbf{Method} & \textbf{Metric} & \multicolumn{3}{c}{\textbf{Town1}} & \multicolumn{3}{c}{\textbf{Town4}} & \multicolumn{3}{c}{\textbf{Town7}} & \multicolumn{3}{c}{\textbf{TOWN10}} \\
\cmidrule(lr){3-5}\cmidrule(lr){6-8}\cmidrule(lr){9-11}\cmidrule(lr){12-14}
 &  & \textbf{\emph{p}-values} & \textbf{95\% CI} & \textbf{Effect} & \textbf{\emph{p}-values} & \textbf{95\% CI} & \textbf{Effect} & \textbf{\emph{p}-values} & \textbf{95\% CI} & \textbf{Effect} & \textbf{\emph{p}-values} & \textbf{95\% CI} & \textbf{Effect} \\
\midrule

\multirow{4}{*}{MOSAT} & Violation Rate     & 0.0222 & $[0.0100, 0.0637]$ & 2.185  & 0.0363 & $[0.0032, 0.0493]$ & 1.810 & 0.0175 & $[0.0204, 0.1021]$ & 2.383 & 0.1751 & $[-0.0195, 0.0682]$ & 0.884 \\
 & TOP-K              & 0.0103 & $[-15.51, -4.49]$ & -2.887 & 0.0902 & $[-18.31, 2.31]$  & -1.234 & 0.1163 & $[-45.39, 8.39]$  & -1.095 & 0.0886 & $[-17.09, 2.09]$  & -1.244 \\
 & Parameter Distance & 0.0000 & $[0.0219, 0.0306]$ & 13.688 & 0.0003 & $[0.0251, 0.0390]$ & 6.044 & 0.0108 & $[0.0116, 0.0484]$ & 2.781 & 0.0172 & $[0.0075, 0.0350]$ & 2.442 \\
 & Map Coverage       & 0.0002 & $[3.21, 6.39]$     & 6.383  & 0.0009 & $[2.06, 3.74]$     & 5.306 & 0.0119 & $[1.66, 9.64]$     & 2.719  & 0.0169 & $[0.85, 4.88]$     & 2.462 \\
\hline

\multirow{4}{*}{GARL} & Violation Rate     & 0.0046 & $[0.0331, 0.0669]$ & 4.141 & 0.0116 & $[0.0134, 0.0454]$ & 2.890 & 0.0295 & $[0.0086, 0.0826]$ & 2.067 & 0.3211 & $[-0.0186, 0.0551]$ & 0.585 \\
 & TOP-K              & 0.0171 & $[-18.01, -1.99]$ & -2.429 & 0.0187 & $[-18.76, -2.74]$ & -2.374 & 0.0359 & $[-45.06, -2.94]$ & -1.905 & 0.0076 & $[-23.52, -7.48]$ & -3.347 \\
 & Parameter Distance & 0.0002 & $[0.0230, 0.0330]$ & 8.686 & 0.0001 & $[0.0292, 0.0413]$ & 8.043 & 0.0120 & $[0.0097, 0.0418]$ & 2.673 & 0.0306 & $[0.0035, 0.0355]$ & 2.055 \\
 & Map Coverage       & 0.0004 & $[4.03, 7.87]$     & 5.908 & 0.0084 & $[0.47, 2.82]$     & 3.050 & 0.0001 & $[8.28, 12.74]$    & 9.252 & 0.0015 & $[2.48, 6.11]$     & 4.945 \\
\hline

\multirow{4}{*}{KING} & Violation Rate     & 0.2793 & $[-0.0233, 0.0095]$ & -0.622 & 0.0040 & $[0.0223, 0.0502]$ & 3.509 & 0.1010 & $[-0.0154, 0.0936]$ & 1.140 & 0.0349 & $[-0.0019, 0.0669]$ & 1.669 \\
 & TOP-K              & 0.0724 & $[-12.82, 1.82]$ & -1.466 & 0.0031 & $[-19.20, -5.30]$ & -3.736 & 0.4248 & $[-22.59, 10.59]$ & -0.461 & 0.0114 & $[-21.85, -4.65]$ & -2.778 \\
 & Parameter Distance & 0.0000 & $[0.0270, 0.0370]$ & 14.000 & 0.0004 & $[0.0230, 0.0370]$ & 5.886 & 0.0106 & $[0.0116, 0.0504]$ & 2.809 & 0.0225 & $[0.0050, 0.0438]$ & 2.257 \\
 & Map Coverage       & 0.0008 & $[3.61, 5.40]$     & 5.158 & 0.0028 & $[1.39, 3.39]$     & 3.888 & 0.1351 & $[-1.33, 9.34]$    & 0.960 & 0.0268 & $[0.32, 5.64]$     & 1.910 \\

\bottomrule
\end{tabular}}
\label{ttest}
\end{table}

\begin{figure}[!t]
    \centering
    \begin{minipage}{0.32\linewidth}
        \centering
        \includegraphics[width=\linewidth]{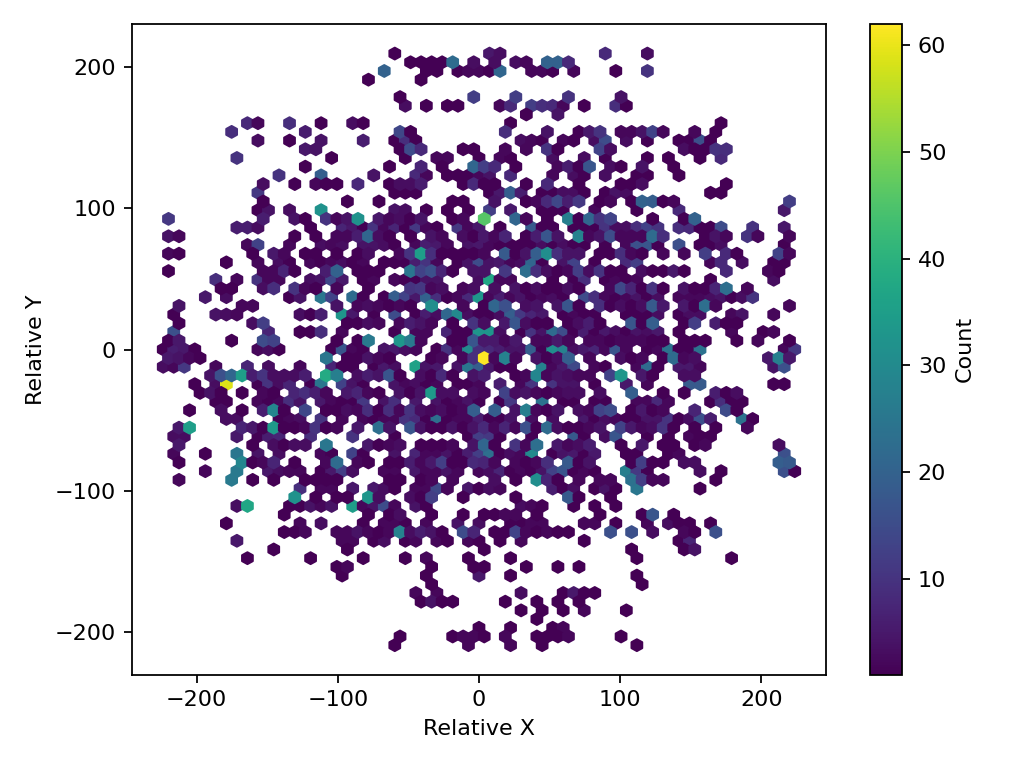}
        \subcaption{GA in Town 10}
    \end{minipage}
    \hfill
    \begin{minipage}{0.32\linewidth}
        \centering
        \includegraphics[width=\linewidth]{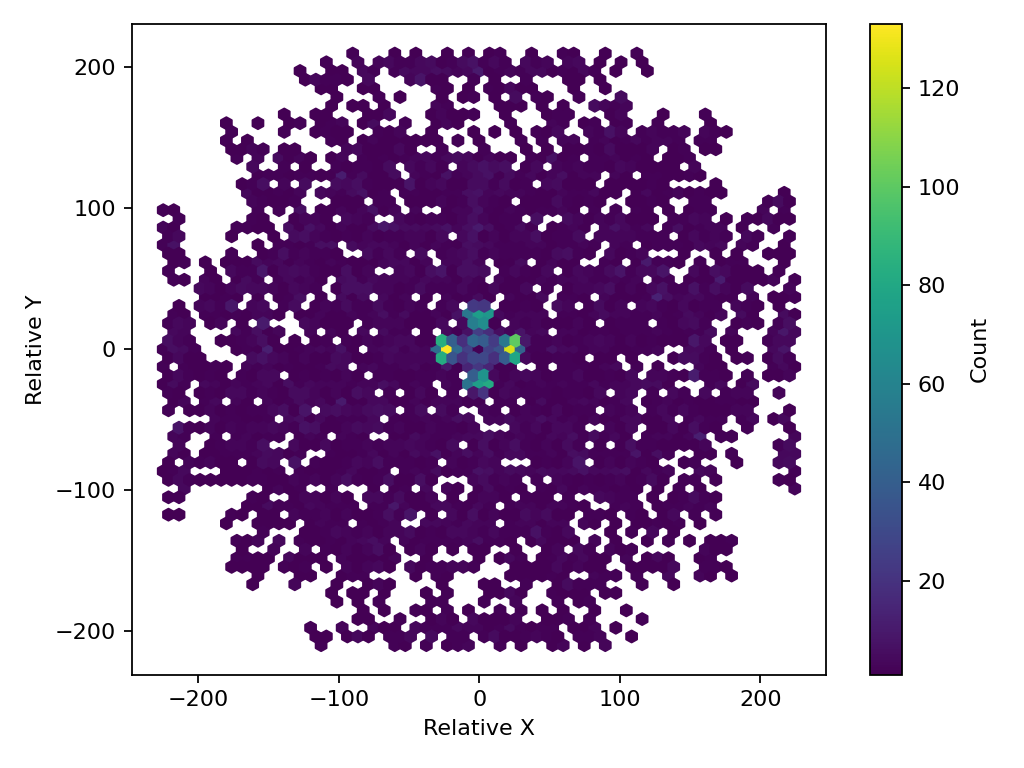}
        \subcaption{ARSG+SVGD in Town 10}
    \end{minipage}
     \hfill
    \begin{minipage}{0.32\linewidth}
        \centering
        \includegraphics[width=\linewidth]{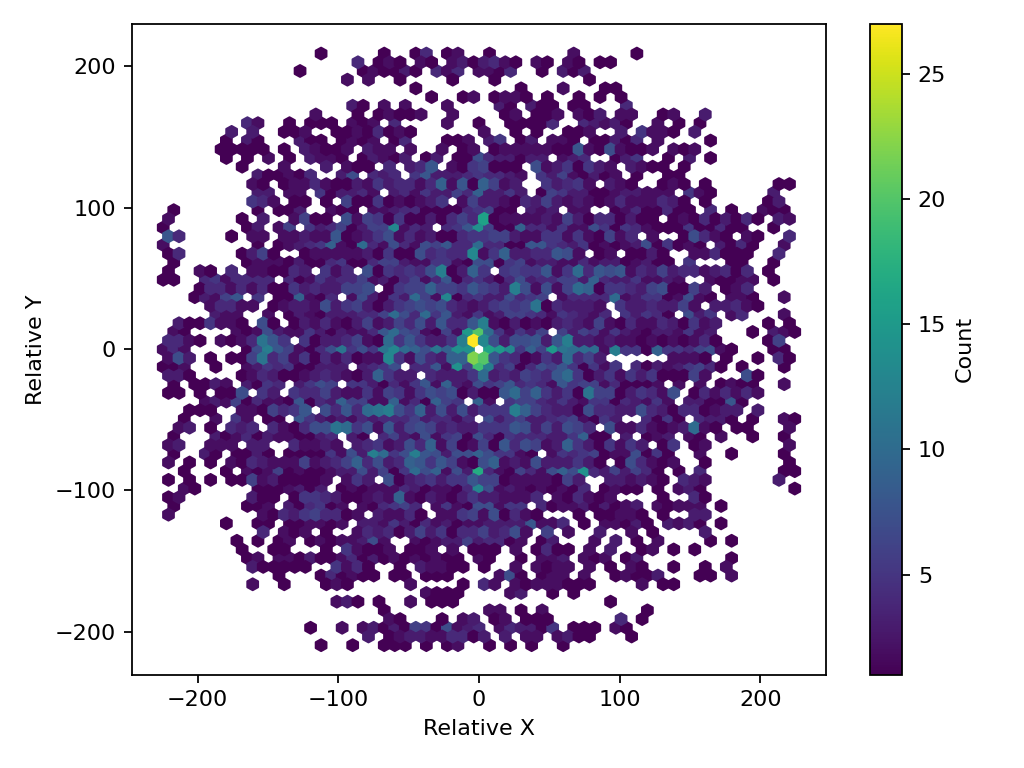}
        \subcaption{Random in Town 10}
    \end{minipage}

    \begin{minipage}{0.32\linewidth}
        \centering
        \includegraphics[width=\linewidth]{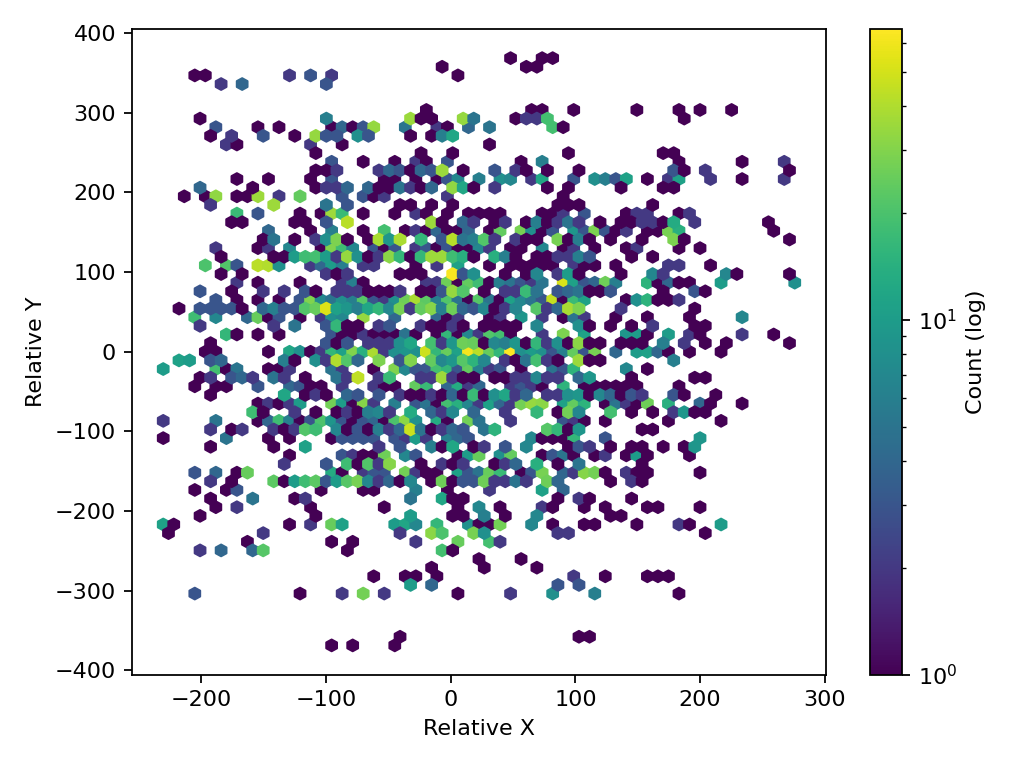}
        \subcaption{GA in Town 07}
    \end{minipage}
    \hfill
    \begin{minipage}{0.32\linewidth}
        \centering
        \includegraphics[width=\linewidth]{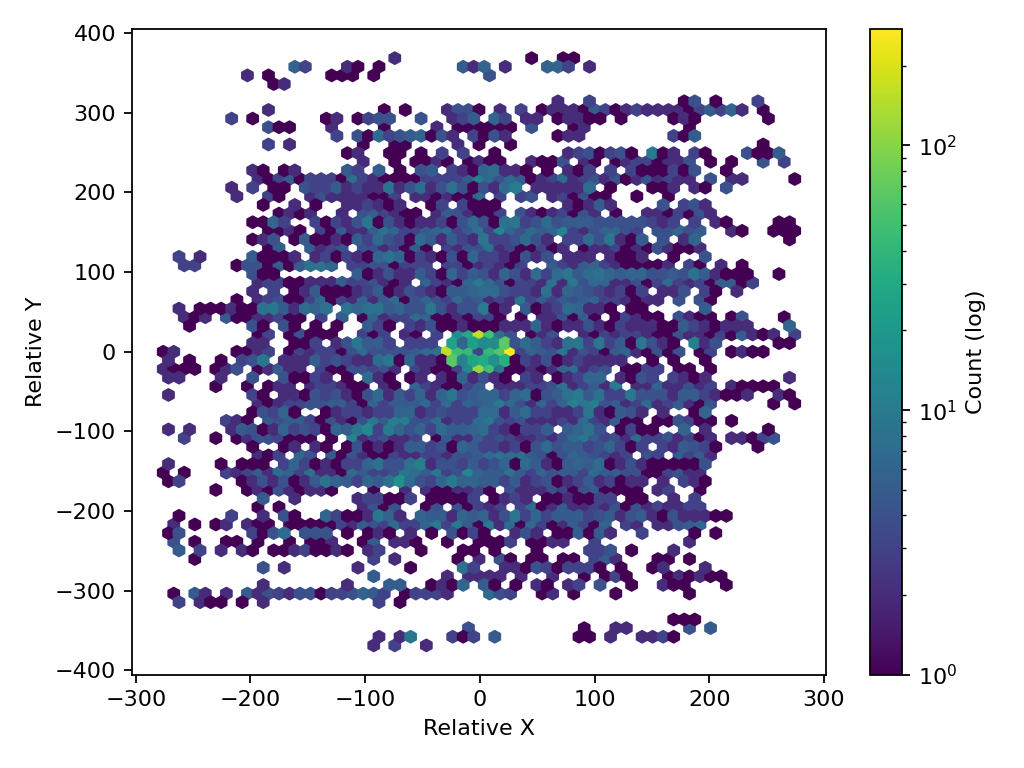}
        \subcaption{ARSG+SVGD in Town 07}
    \end{minipage}
     \hfill
    \begin{minipage}{0.32\linewidth}
        \centering
        \includegraphics[width=\linewidth]{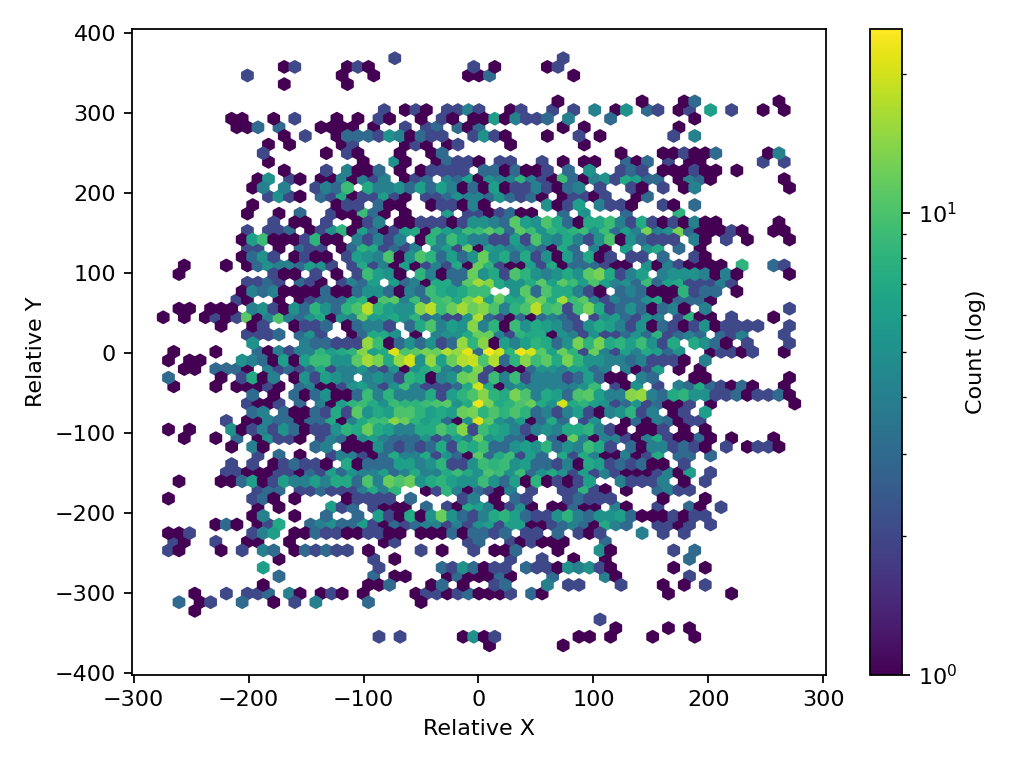}
        \subcaption{Random in Town 07}
    \end{minipage}

    \begin{minipage}{0.32\linewidth}
        \centering
        \includegraphics[width=\linewidth]{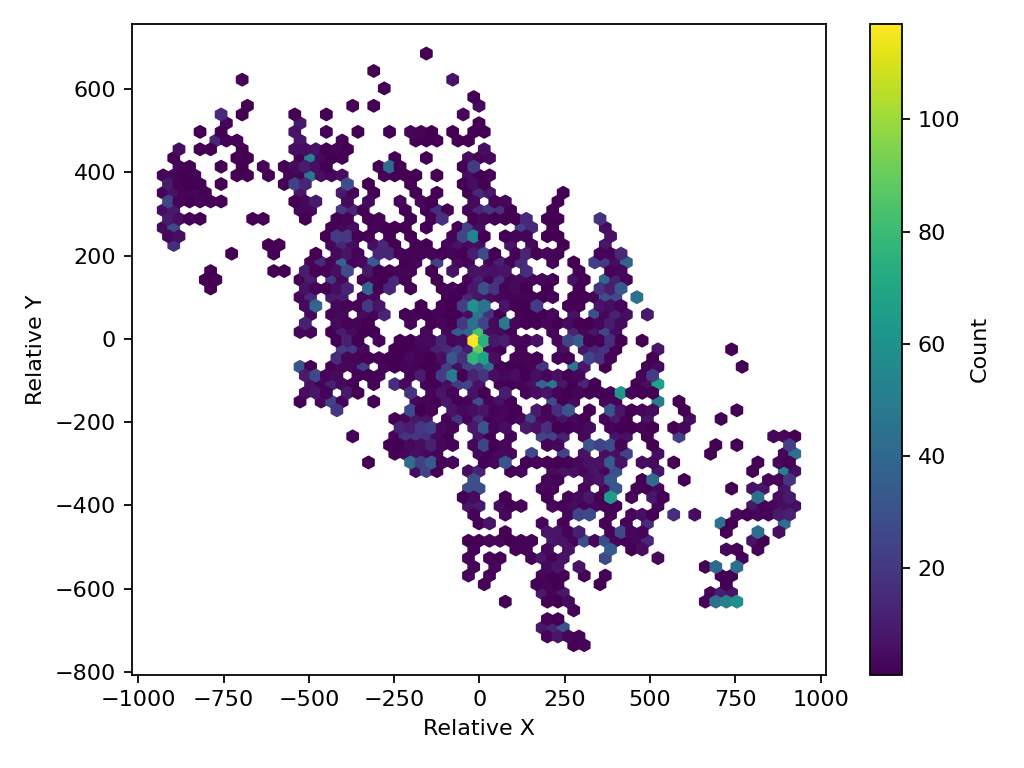}
        \subcaption{GA in Town 04}
    \end{minipage}
    \hfill
    \begin{minipage}{0.32\linewidth}
        \centering
        \includegraphics[width=\linewidth]{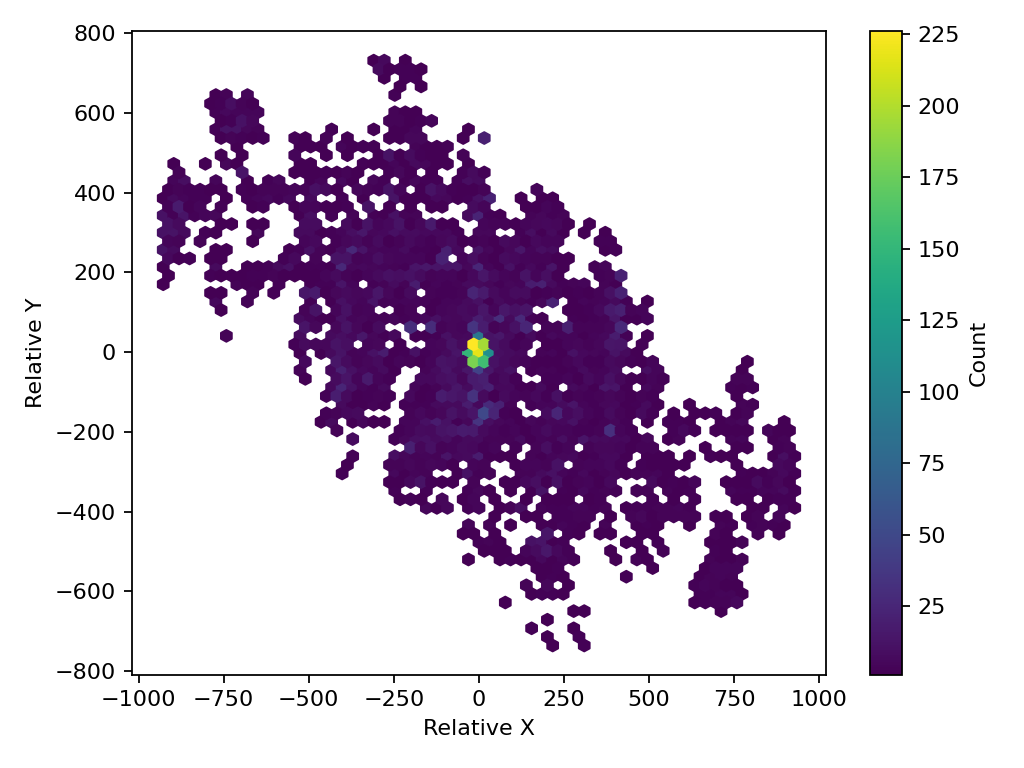}
        \subcaption{ARSG+SVGD in Town 04}
    \end{minipage}
     \hfill
    \begin{minipage}{0.32\linewidth}
        \centering
        \includegraphics[width=\linewidth]{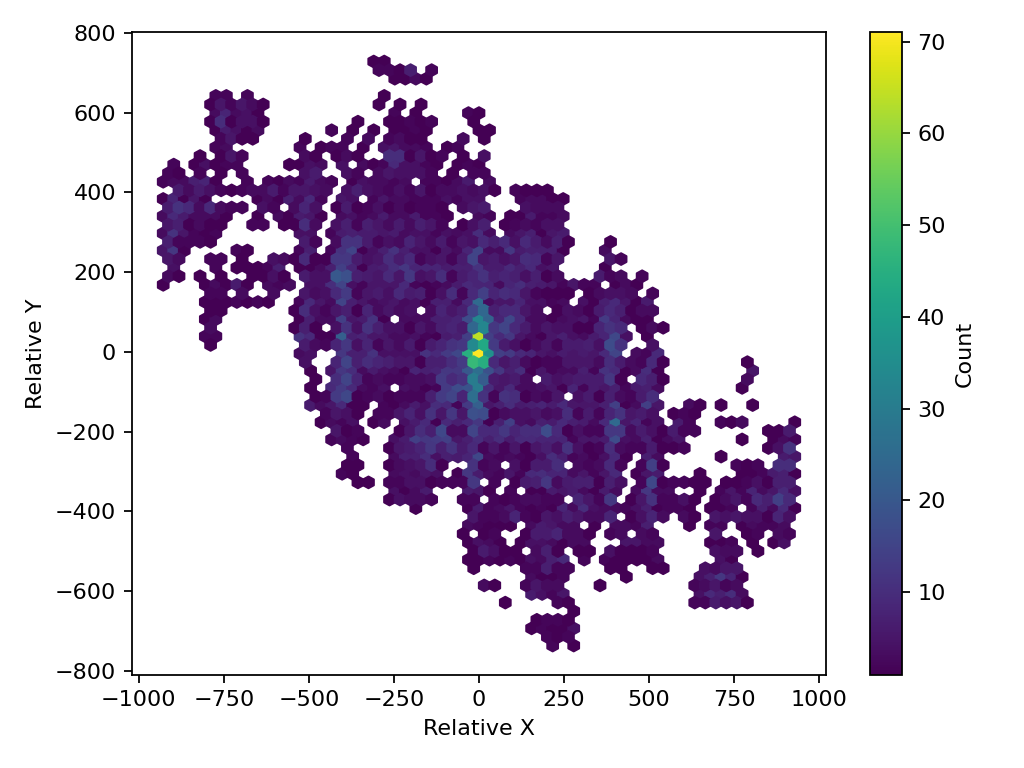}
        \subcaption{Random in Town 04}
    \end{minipage}

\begin{minipage}{0.32\linewidth}
        \centering
        \includegraphics[width=\linewidth]{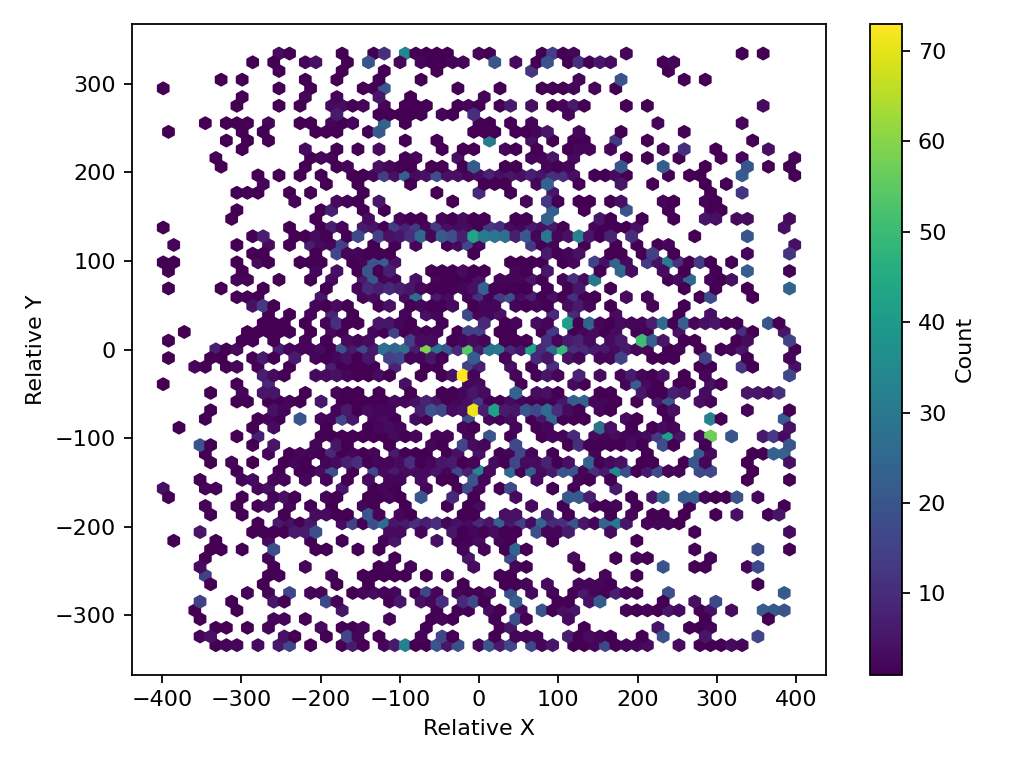}
        \subcaption{GA in Town 01}
    \end{minipage}
    \hfill
    \begin{minipage}{0.32\linewidth}
        \centering
        \includegraphics[width=\linewidth]{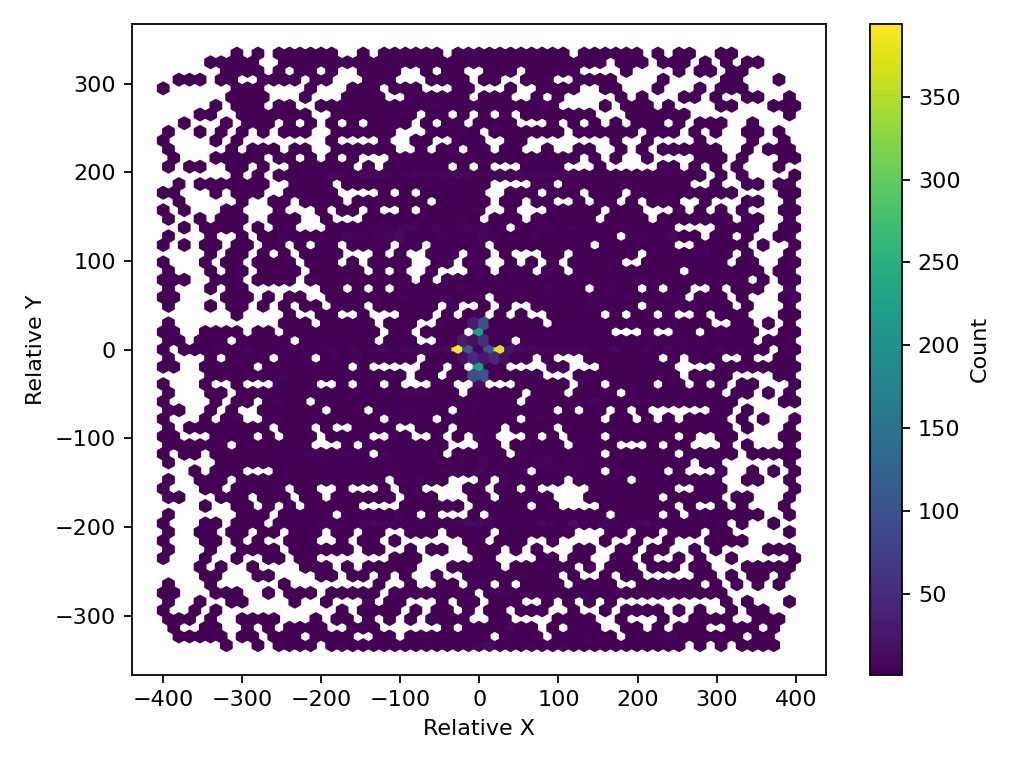}
        \subcaption{ARSG+SVGD in Town 01}
    \end{minipage}
     \hfill
    \begin{minipage}{0.32\linewidth}
        \centering
        \includegraphics[width=\linewidth]{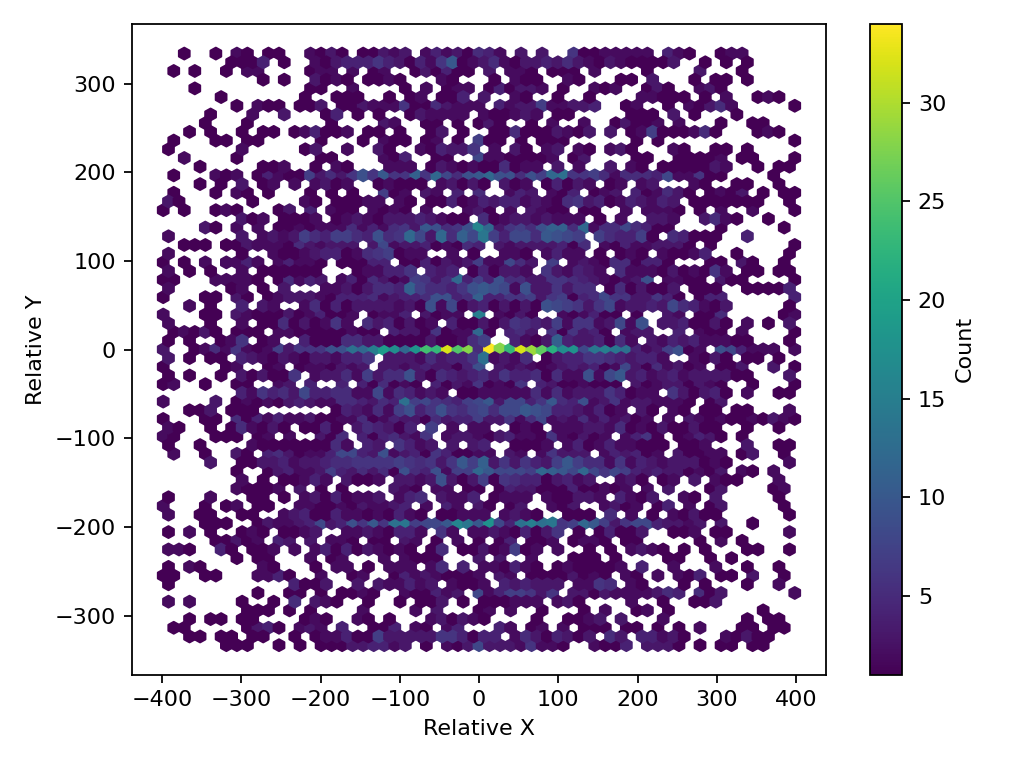}
        \subcaption{Random in Town 01}
    \end{minipage}
    
    \caption{Scatter plot of the initial relative position between ego vehicle and dynamic objects across different maps generated by different offline methods. }
    \label{dis1}
\end{figure}

\begin{figure}[!t]
    \centering
    \begin{minipage}{0.32\linewidth}
        \centering
        \includegraphics[width=\linewidth]{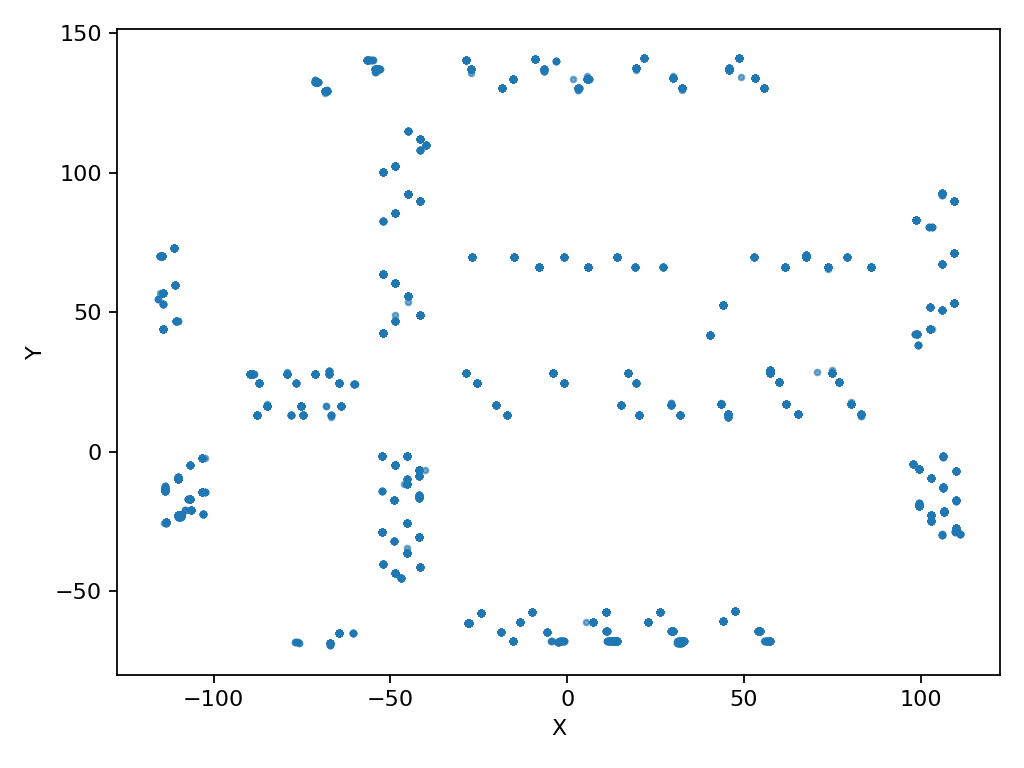}
        \subcaption{GA in Town 10}
    \end{minipage}
    \hfill
    \begin{minipage}{0.32\linewidth}
        \centering
        \includegraphics[width=\linewidth]{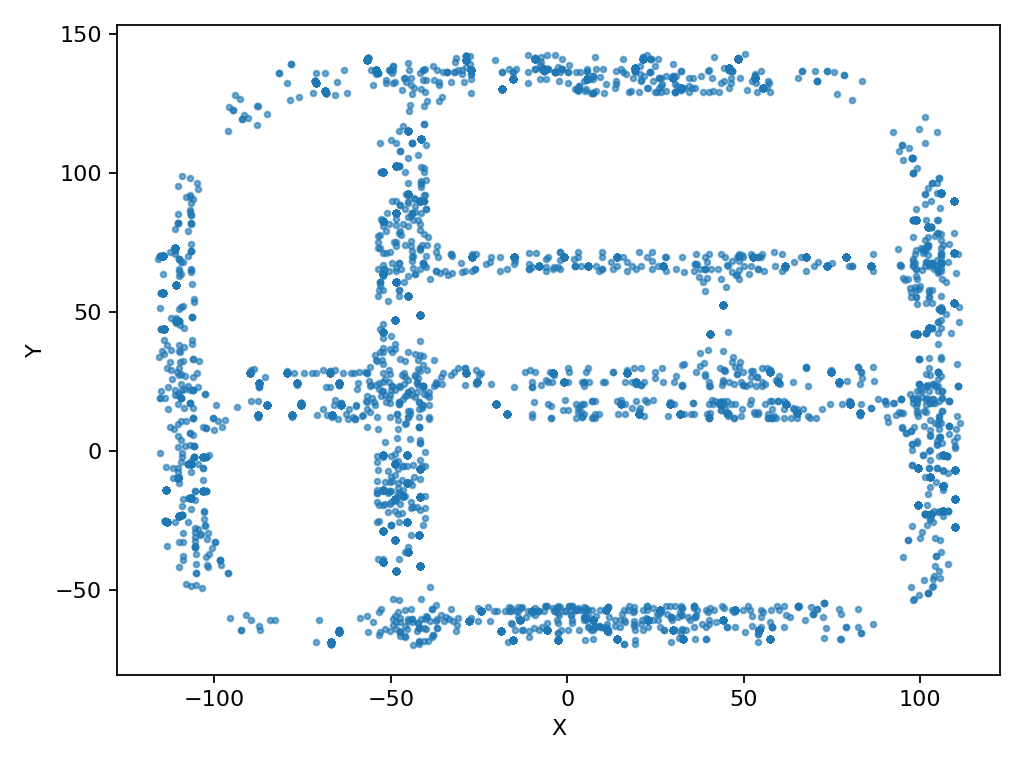}
        \subcaption{ARSG+SVGD in Town 10}
    \end{minipage}
     \hfill
    \begin{minipage}{0.32\linewidth}
        \centering
        \includegraphics[width=\linewidth]{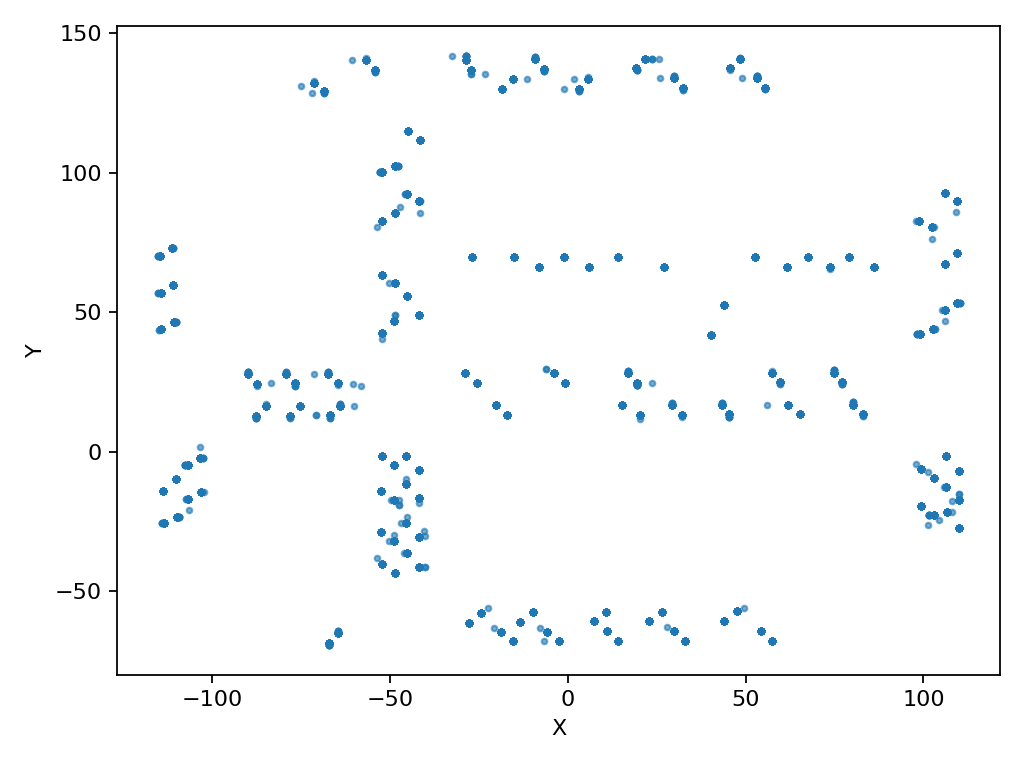}
        \subcaption{Random in Town 10}
    \end{minipage}

    \begin{minipage}{0.32\linewidth}
        \centering
        \includegraphics[width=\linewidth]{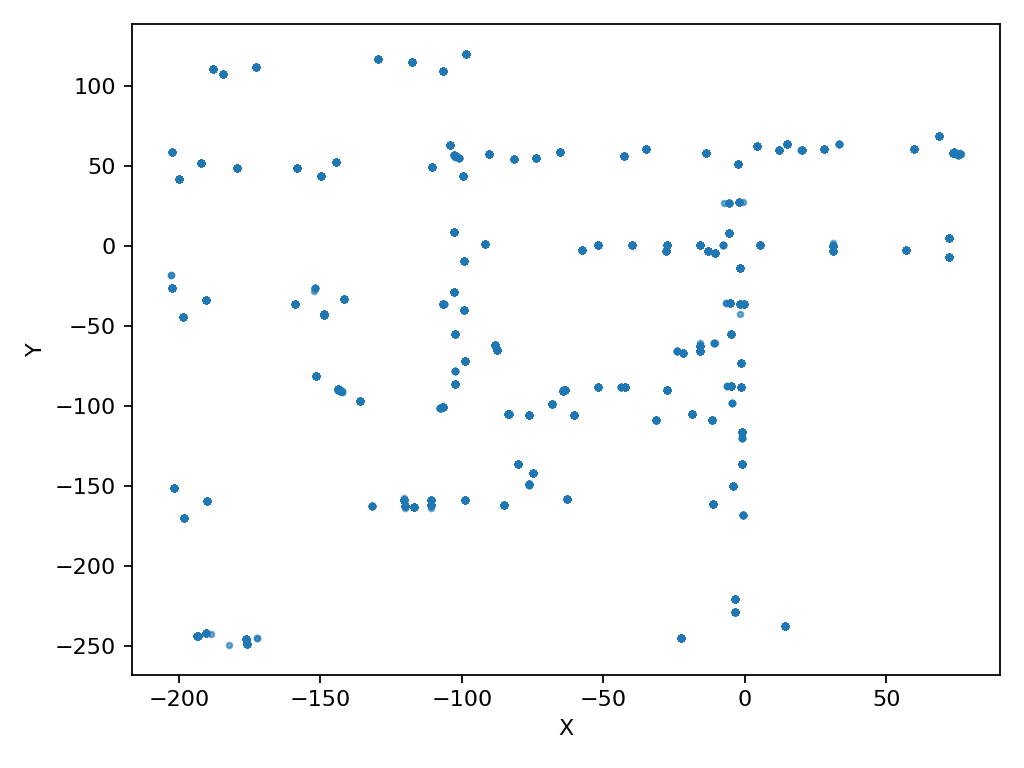}
        \subcaption{GA in Town 07}
    \end{minipage}
    \hfill
    \begin{minipage}{0.32\linewidth}
        \centering
        \includegraphics[width=\linewidth]{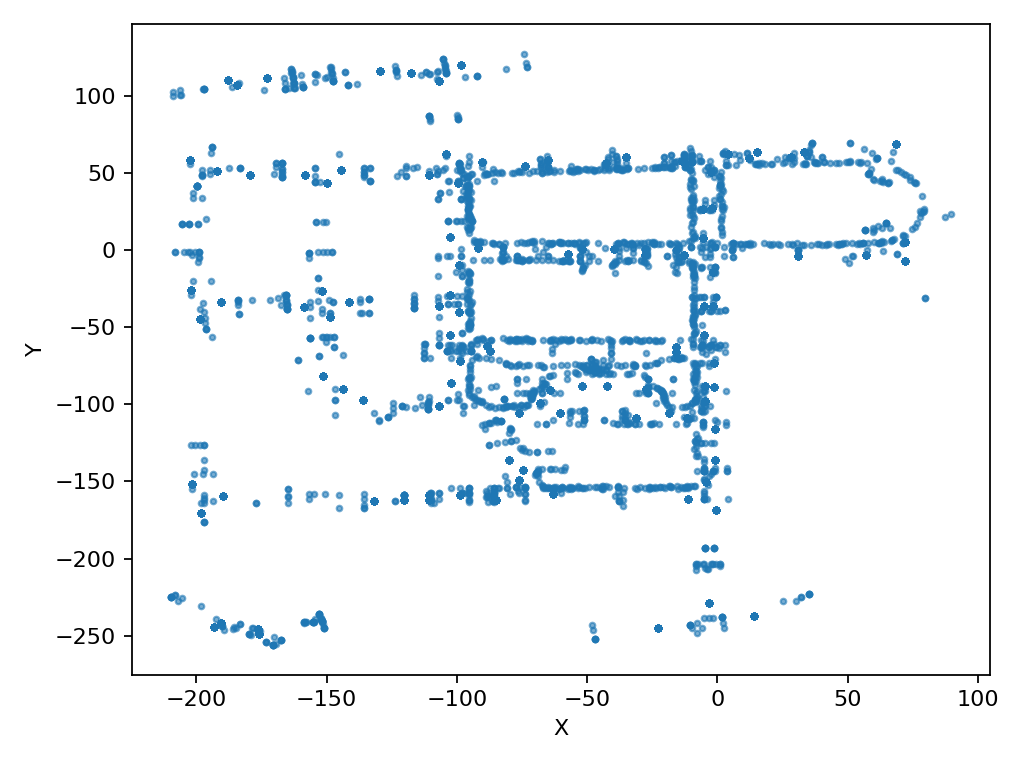}
        \subcaption{ARSG+SVGD in Town 07}
    \end{minipage}
     \hfill
    \begin{minipage}{0.32\linewidth}
        \centering
        \includegraphics[width=\linewidth]{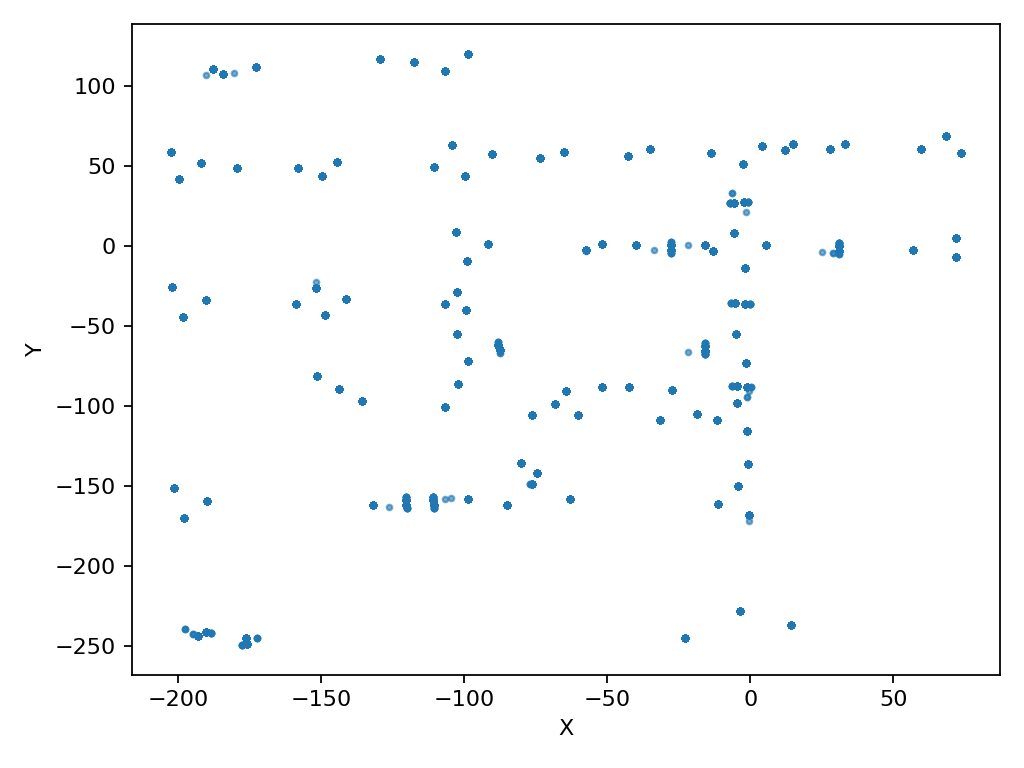}
        \subcaption{Random in Town 07}
    \end{minipage}

     \begin{minipage}{0.32\linewidth}
        \centering
        \includegraphics[width=\linewidth]{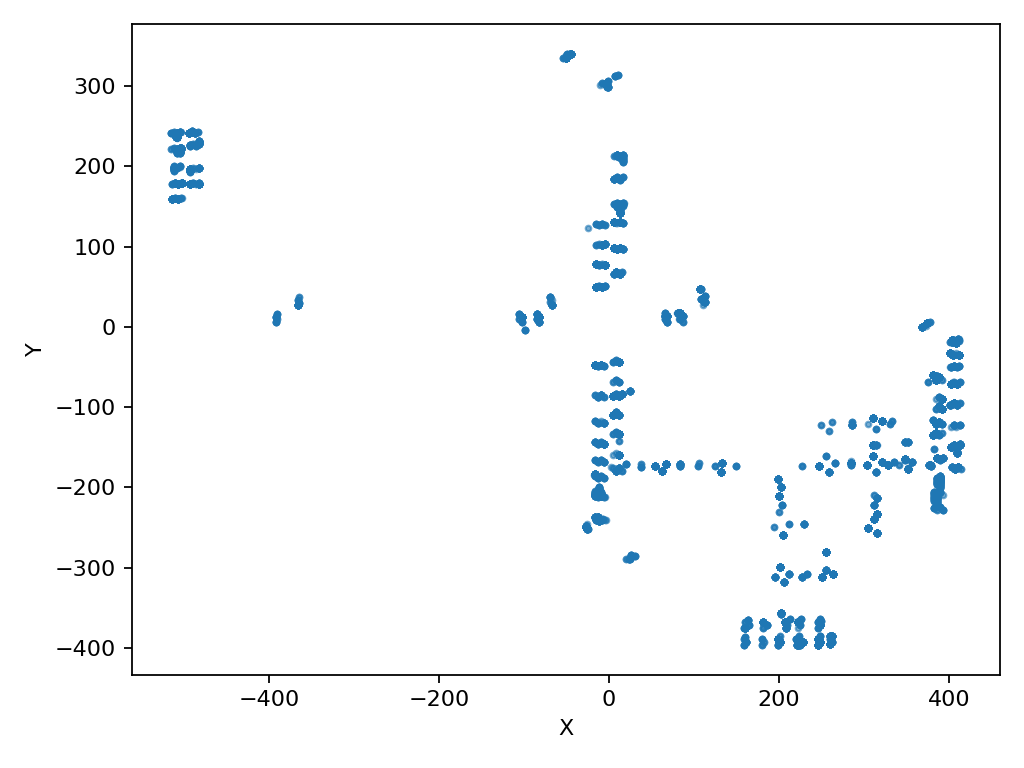}
        \subcaption{GA in Town 04}
    \end{minipage}
    \hfill
    \begin{minipage}{0.32\linewidth}
        \centering
        \includegraphics[width=\linewidth]{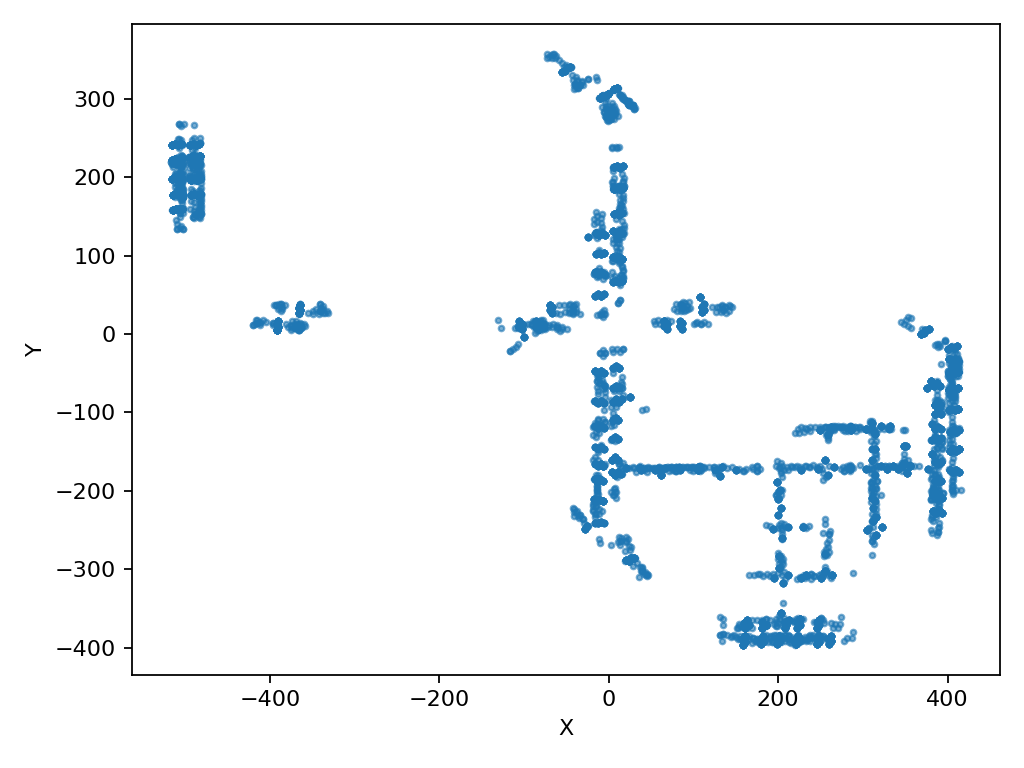}
        \subcaption{ARSG+SVGD in Town 04}
    \end{minipage}
     \hfill
    \begin{minipage}{0.32\linewidth}
        \centering
        \includegraphics[width=\linewidth]{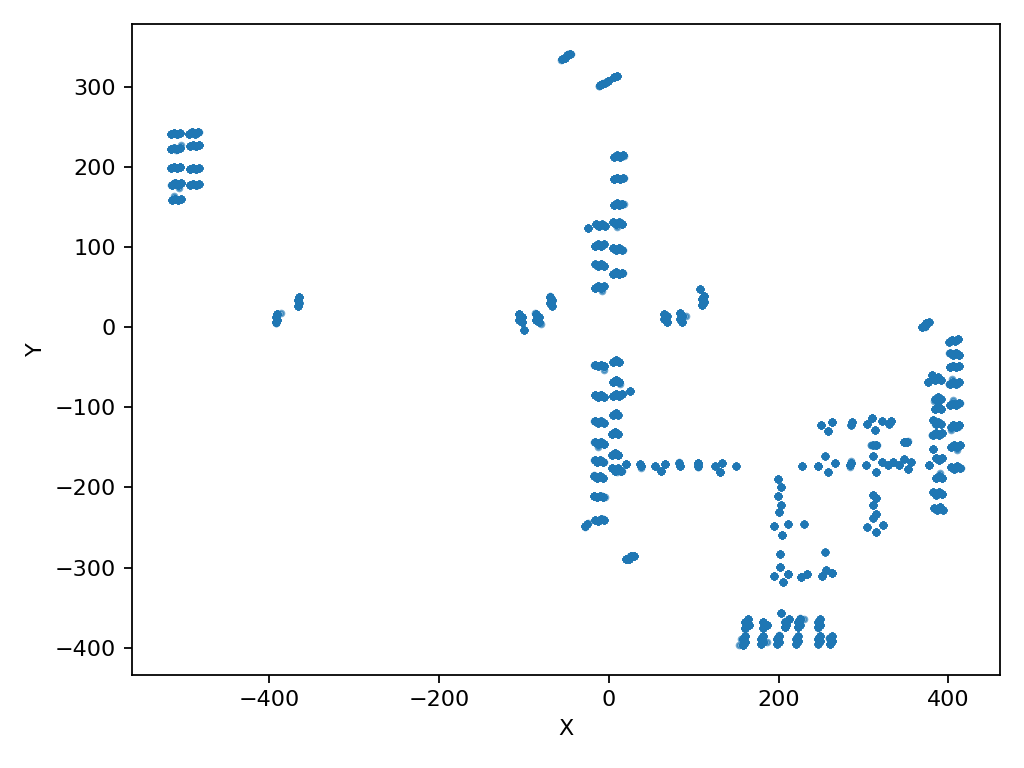}
        \subcaption{Random in Town 04}
    \end{minipage}

     \begin{minipage}{0.32\linewidth}
        \centering
        \includegraphics[width=\linewidth]{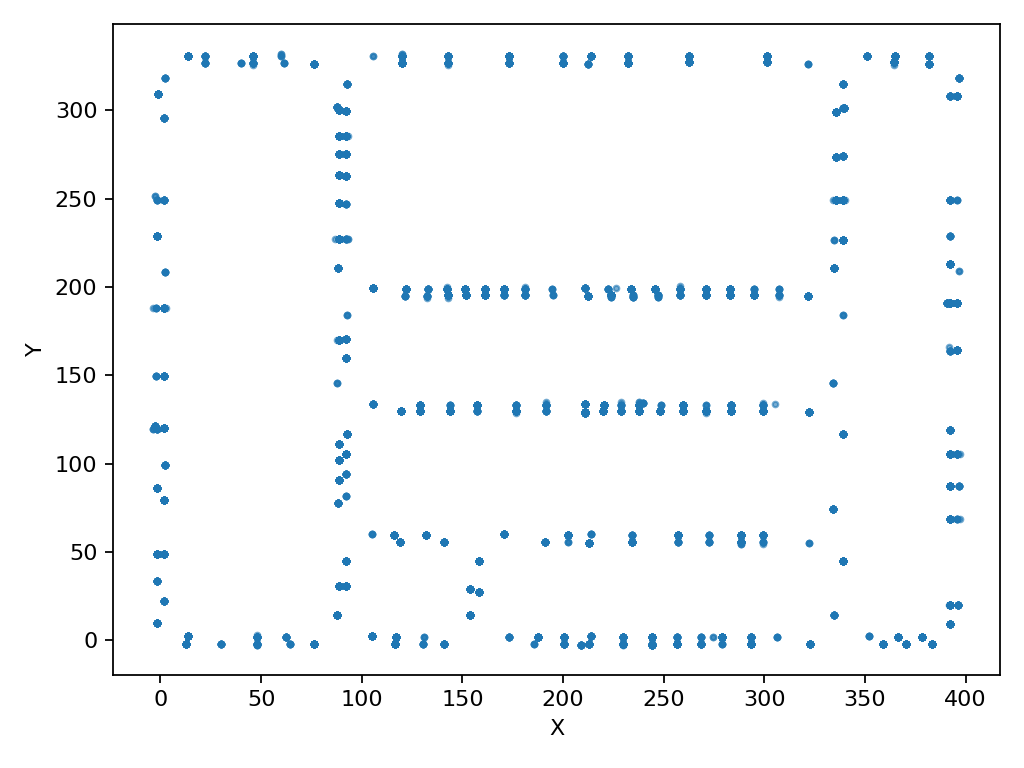}
        \subcaption{GA in Town 01}
    \end{minipage}
    \hfill
    \begin{minipage}{0.32\linewidth}
        \centering
        \includegraphics[width=\linewidth]{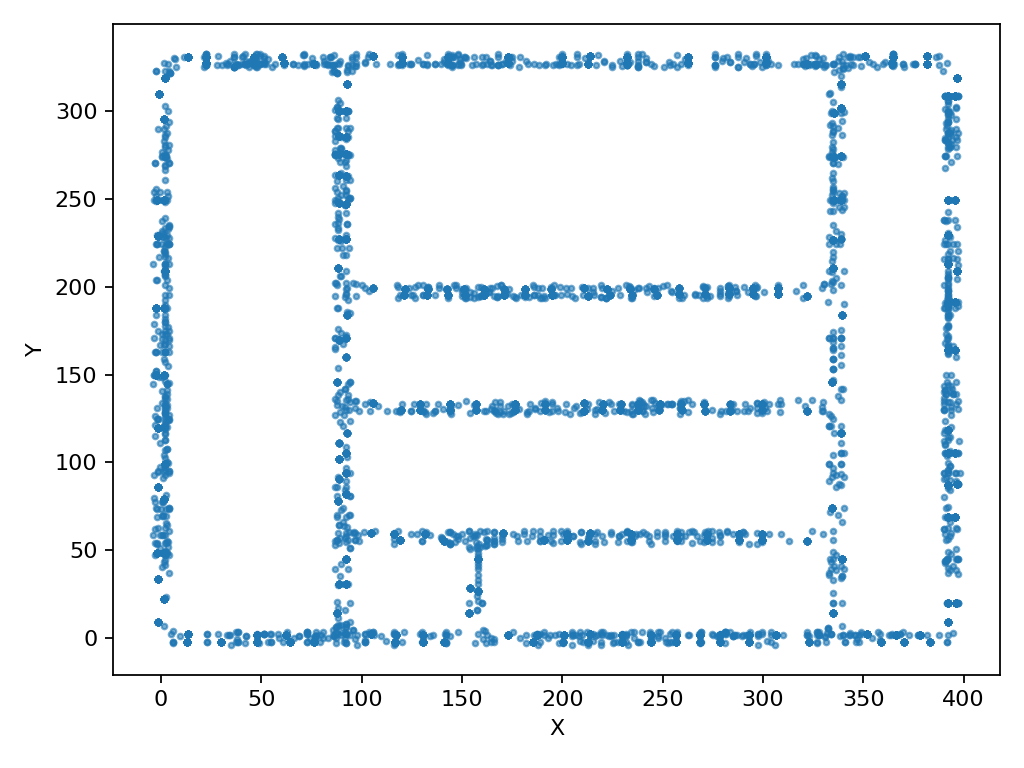}
        \subcaption{ARSG+SVGD in Town 01}
    \end{minipage}
     \hfill
    \begin{minipage}{0.32\linewidth}
        \centering
        \includegraphics[width=\linewidth]{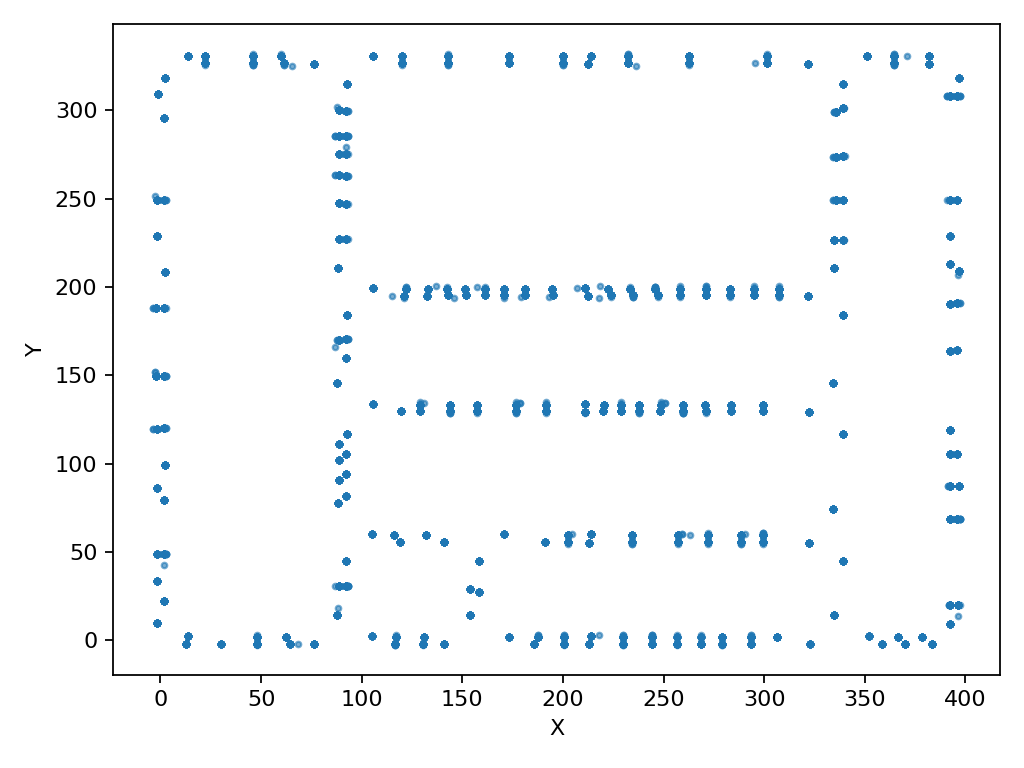}
        \subcaption{Random in Town 01}
    \end{minipage}
    
    \caption{Scatter plot of the absolute initial position of dynamic objects across different maps generated by different offline methods. }
    \label{dis2}
\end{figure}

% \begin{figure}[!t]
%     \centering
%     \begin{minipage}{0.49\linewidth}
%         \centering
%         \includegraphics[width=\linewidth]{img/histo/overlay_rel_x_hist.png}
%         \subcaption{Example of Video 3 scenario}
%     \end{minipage}
%     \hfill
%     \begin{minipage}{0.49\linewidth}
%         \centering
%         \includegraphics[width=\linewidth]{img/histo/overlay_rel_y_hist.png}
%         \subcaption{Example of Video 12 scenario}
%     \end{minipage}

%     \caption{Outlier analysis}
%     \label{out1}
% \end{figure}

Table \ref{tab:exp2_baseline} reports quantitative results comparing \tool\ with baseline methods across multiple Apollo maps. Overall, \tool\ consistently outperforms all baselines in \textbf{Safety Violation Rate} (by up to \textbf{27.68\%}), \textbf{TOP-10} (by up to \textbf{24.04\%}), \textbf{parameter distance} (by up to \textbf{9.60\%}), and \textbf{map coverage} (by up to \textbf{16.78\%}), while incurring no additional runtime overhead. Regarding trajectory coverage, performance mainly depends on the online control component. Due to the gradient-based controller used in KING and its continuous action space, exploration tends to be broader than that of RL-based and GA-based control methods. We then conduct significant tests on the experimental results.

% In \textit{Town1}, \tool\ does not significantly outperform baselines and slightly underperforms GARL in coverage metrics. This is because \textit{Town4} and \textit{Town10}, with four and two-lane roads respectively, provide greater search flexibility, benefiting MARL and ART. In contrast, \textit{Town1} has only single-lane roads, limiting MARL and ART effectiveness due to our constrained action space for MARL training efficiency, which restricts surrounding vehicles from changing lanes. 

% From Table \ref{ttest}, we find that \tool\ differs significantly from other methods on most metrics; in particular, for \textbf{parameter distance} and \textbf{map coverage}—the strengths of \tool—\tool\ consistently shows significant improvements over all baselines. This superior performance primarily stems from the diverse, failure-inducing initial scenarios generated by our approach. 

From Table \ref{ttest}, we observe that \tool\ differs significantly from other methods on most metrics, as indicated by the paired two-sided \emph{t}-tests. 
In particular, for \textbf{parameter distance} and \textbf{map coverage}—the strengths of \tool—\tool\ consistently achieves statistically significant improvements over all baselines, with 95\% confidence intervals largely excluding zero and generally large effect sizes. 

To validate the assumptions of the \emph{t}-test, we conducted the Shapiro–Wilk test on the paired differences, which showed no significant deviations from normality (all $p > 0.05$). 
Given the small sample size ($n=4$), we further applied the non-parametric Wilcoxon signed-rank test. While the Wilcoxon test does not always reach statistical significance (e.g., $p=0.125$ in several cases), this is expected due to its limited statistical power with small samples. Importantly, the observed performance trends remain consistent with those of the paired \emph{t}-test, and the effect sizes remain large.

This superior performance stems from the diverse, failure-inducing initial scenarios generated by our approach, enabling broader parameter space exploration and improved coverage; additional cross-town and cross-agent significance tests are reported in \href{https://github.com/lfeng0722/PtoP/blob/main/FSE_2026_Supplementary.pdf}{\textbf{Tables~1--2 in the supplementary material}}. To substantiate this, Figures~\ref{dis1} and \ref{dis2} show (i) scatter plots of initial \emph{relative} positions between the ego vehicle and dynamic objects across maps generated by different offline methods, and (ii) scatter plots of the \emph{absolute} initial positions of dynamic objects.
%This superior performance primarily stems from the diverse, failure-inducing initial scenarios generated by our approach, which enable broader exploration of the parameter space and improved coverage. A further cross-town and cross-agent significant tests are reported in \href{https://anonymous.4open.science/r/PtoP-75B3/FSE_2026_Supplementary.pdf}{\textbf{the Table 1 and 2 of supplementary material}}.

%\par To substantiate this, Figures \ref{dis1} and \ref{dis2} present (i) scatter plots of the initial \emph{relative} positions between the ego vehicle and dynamic objects across different maps generated by different offline methods, and (ii) scatter plots of the \emph{absolute} initial positions of dynamic objects across those maps.

For the relative position, which serves as a proxy for efficiency, the distribution is expected to be approximately normal: intuitively, dynamic objects nearer to the ego vehicle have a higher likelihood of causing a safety violation. From Figure~\ref{dis1}, we observe that ARSG+SVGD concentrates samples near the peak of this distribution, whereas the other baselines are more evenly spread; GA, in particular, clusters around distinct local minima, while Random remains broadly uniform. This behavior arises because SVGD combines an attractive term that pulls particles toward high-hazard modes. In contrast, GA tends to discover multiple local minima and, through its inheritance mechanism, propagates these minima to subsequent generations, which limits efficiency.

For the absolute position, which serves as a proxy for diversity, the samples should be broadly and evenly distributed across the map. Figure~\ref{dis2} shows that ART+SVGD places dynamic objects more evenly and in a denser, more uniform pattern across the map, whereas GA and Random are noticeably sparser. Together with the numerical results, these observations indicate that \tool\ produces more diverse and failure-inducing initial scenarios.

Moreover, regarding the number of safety violations found within 6 hours, although the methodologies differ, under the same evaluation settings our method achieves roughly twice the safety-violation discovery rate of TARGET (91.31 vs.\ 40.5). This result highlights the importance of search-based testing.

\tool is not highly sensitive to SVGD parameters within reasonable ranges. We use 5 particles in all experiments as it provides a good balance between exploration and exploitation; increasing to 20 particles in Town4 reduces parameter distance to 0.25, reflecting weaker exploration under online training. The kernel bandwidth is adaptively computed using the median heuristic, making it self-adjusting across towns and episodes. Temperature is set to 1 by default; lower values (e.g., 0.1) still yield stable behavior (~0.29 distance), whereas very high temperatures (e.g., 5) collapse particles into a single mode and cause unrealistic stacking. Overall, across all reasonable settings, PtoP retains its performance gains. Moreover, we further analyze the failure taxonomy in \href{https://github.com/lfeng0722/PtoP/blob/main/FSE_2026_Supplementary.pdf}{\textbf{the Figure 2 of supplementary material}}.

\subsection{Ablation Study (RQ3)}
% ====================== Ablation ======================
\begin{table}[t]
\centering
\caption{Ablation study across towns.}

\resizebox{\columnwidth}{!}{
\begin{tabular}{l l c c c c c}
\hline
\textbf{Method} & \textbf{Town} & \textbf{Safety Violation \%} & \textbf{Top-10} & \textbf{Parameter distance} & \textbf{Map Coverage \%} & \textbf{Time Consumption (hours)} \\
\hline
\multirow{4}{*}{\textit{\tool \ without ART}} 
 & Town1 & 20.21\% & 50.25 & 0.301 & 29.70\% & 6 \\
 & Town4 & 15.06\%  & 75.75  & 0.292 & 20.67\%  & 6 \\
 & Town7 & 27.19\% & 51 & 0.297 & 70.25\% & 6 \\
 & Town10 & 16.25\% & 50.75 & 0.292 & 65.18\% & 6 \\
\hline
\multirow{4}{*}{\textit{\tool\ with GA}} 
 & Town1 & 20.69\% & 49.25 & 0.3 & 25.51\% & 6 \\
 & Town4 & 14.38\%  & 74.5  & 0.291 & 15.51\%  & 6 \\
 & Town7 & 27.25\% & 51 & 0.283 & 59.15\%  & 6 \\
 & Town10 & 18.51\% & 56 & 0.293 & 56.49\% & 6 \\
\hline
\end{tabular}}
\label{tab:exp2_ablation}
\end{table}

\par Table~\ref{tab:exp2_ablation} presents the ablation study results. For \tool\ with GA, we reach the same conclusion as in RQ2 i.e. GA exhibits limited performance in \textbf{map coverage} and \textbf{diversity}, which is aligned with MOSAT and GARL. For \tool\ without ART, the method still maintains high \textbf{map coverage}, but the \textbf{parameter distance} drops to the level of GA and Random. This aligns with our hypothesis: ART tends to identify additional failure modes, while SVGD explores within each discovered mode. Consequently, removing ART reduces \tool's ability to uncover more failure modes, yet its exploration capability remains, thereby preserving coverage.

\subsection{Fidelity (RQ4)}
\begin{figure}
    \centering
    \includegraphics[width=0.7\linewidth]{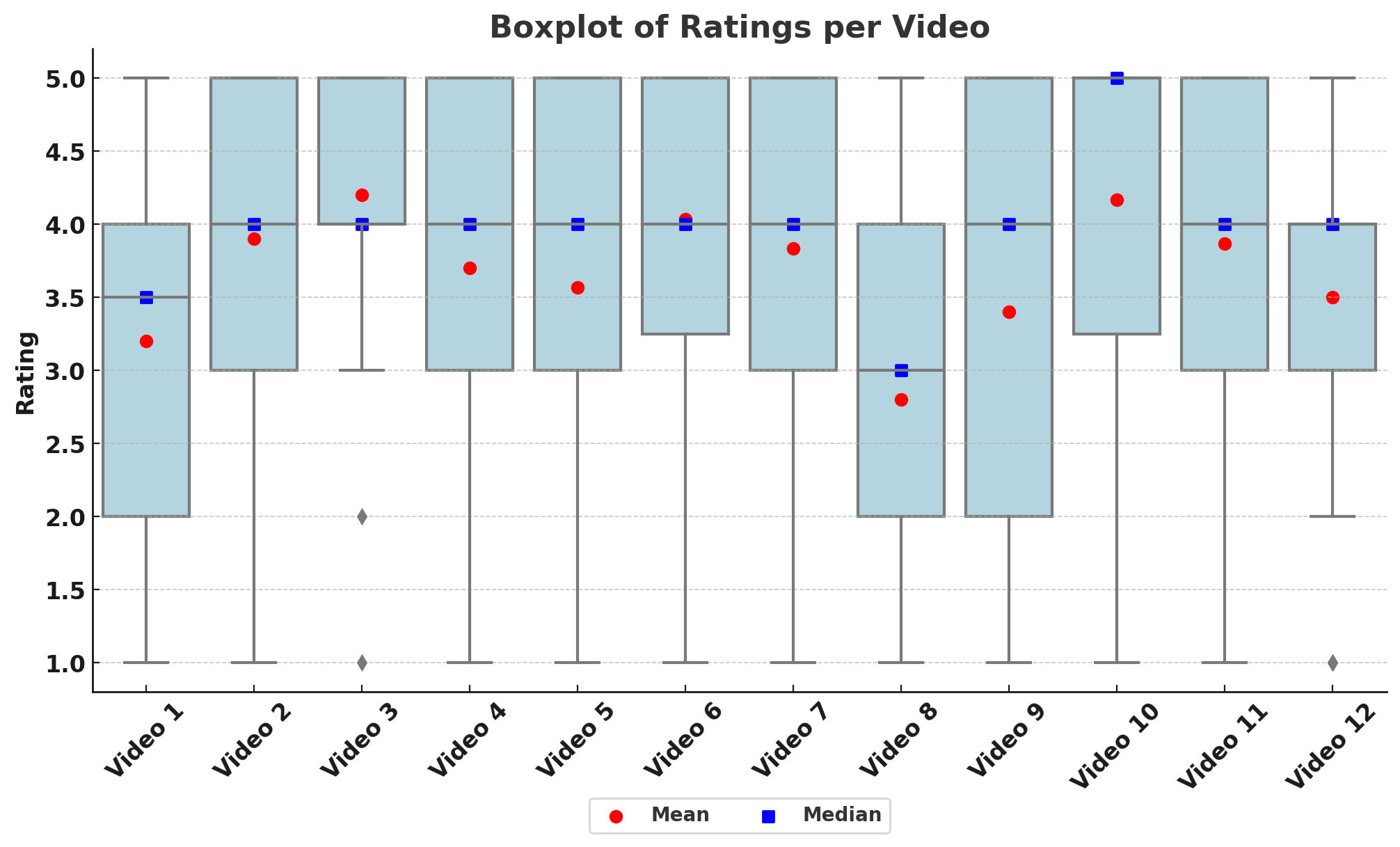}
    \vspace{-3mm}
    \caption{ Boxplots for user ratings on sampled video with outliers.}
    \label{user_study}
\end{figure}
Figure \ref{user_study} presents a box plot of realism ratings from the user study, where the red and blue dots represent the mean and median ratings, respectively. The median rating for all videos exceeds 3, indicating a ``Neutral'' or higher realism perception. Only video 8 has a mean rating below 3. The weighted Fleiss’ Kappa for user ratings is \textbf{0.714}, indicating ``Substantial Agreement.'' However, outliers remain in videos 3 and 12, analyzed in the next section.
%Figure \ref{user_study} presents a box plot of realism ratings from the user study, where the red dot represents the mean rating, and the blue dot represents the median rating for each video. We observe that the median rating for all videos is above 3, indicating a ``Neutral'' or higher perception of realism. Notably, only one video's mean rating—video 8—falls below 3. Additionally, we computed the weighted Fleiss’ Kappa for user ratings, which resulted in a kappa value of \textbf{0.714}, indicating ``Substantial Agreement.'' However, some outliers remain in video 3 and video 12, which we further analyze in the following section.

\subsubsection{Video 3 \& Video 12 Outlier Analysis}
\begin{figure}[!t]
    \centering
    \begin{minipage}{0.49\linewidth}
        \centering
        \includegraphics[width=\linewidth]{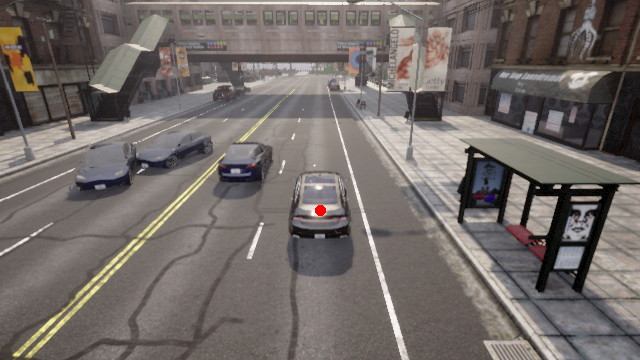}
        \subcaption{Example of Video 3 scenario}
    \end{minipage}
    \hfill
    \begin{minipage}{0.49\linewidth}
        \centering
        \includegraphics[width=\linewidth]{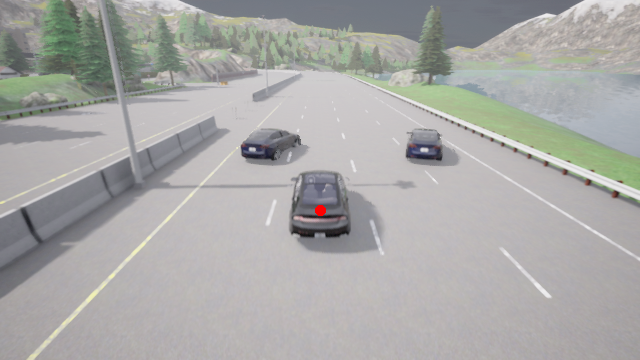}
        \subcaption{Example of Video 12 scenario}
    \end{minipage}

    \caption{Outlier analysis.}
    \label{out1}
\end{figure}

In Video~3, we observe two outliers. One comment states, ``I don't think this would happen in the real world''. This likely refers to the left image of Figure~\ref{out1}, where two surrounding vehicles collide due to concurrent lane changes. In Video~12, one outlier comment is ``They were constantly changing lanes''. The right image of Figure~\ref{out1} shows vehicles attempting multiple lane changes within a short period. Currently, our framework relies on existing online methods; future work will develop stronger online components and incorporate additional regulations (e.g., rule-based constraints) to improve realism.
\section{Threats to Validity}\label{tv}

% The soundness of our approach is subject to threats to validity, which we discuss along the dimensions of internal, construct, and external validity.

\noindent\textbf{Internal Validity.} A key threat lies in the choice of online testing methods. RL-based testers~\cite{liang2023rlaga} suffer from \emph{action realism} issues, where generated behaviors may diverge from plausible driving maneuvers, while gradient-based testers~\cite{hanselmann2022king} face the \emph{catch-on-collision} problem, where optimization halts once a collision occurs. These limitations may affect the extent to which discovered violations reflect realistic driving scenarios. Nevertheless, our framework is \emph{plug-and-play} and integrates these state-of-the-art online testing methods to ensure a fair comparison; developing more robust online testers is orthogonal and left as future work.   

\noindent\textbf{Construct Validity.} The validity of our evaluation also depends on the chosen baselines. To reduce bias, we emphasize the complementary roles of the offline seed generator and the online tester through a hazard learning model, and we compare against both recent offline generators and leading online testers. This setup increases confidence that performance gains arise from our design rather than from limited baseline coverage.

\noindent\textbf{External Validity.}
Although CARLA-based simulation cannot fully replicate real-world sensing noise and long-tail edge cases, our design mitigates the external validity by demonstrating that observed effects persist across diverse road structures and autonomy architectures rather than emerging from a single simulator configuration or system implementation. Concretely, the selected CARLA maps span compact urban (Town1), metropolitan multi-intersection (Town10), highway-dominant (Town4), and rural navigation-complex (Town7) scenarios, thereby covering heterogeneous road topologies, traffic regulations, speed regimes, and interaction patterns. This reduces the risk that results are tied to a single driving context. In parallel, we evaluate three architecturally distinct ADSs—Apollo 8.0 and Autoware as full-stack, industry-scale open-source systems with different planning and control designs, and Traffic Manager as a deterministic, low-complexity baseline—thereby limiting conclusions to neither a specific autonomy stack nor a particular perception–planning–control coupling. 
ISO~34502 and ISO~21448 (SOTIF) are not threats but complementary contexts: our tool offers \emph{exploratory} stress testing for early-stage violation discovery, aligned with SOTIF’s focus on performance limitations under nominal conditions.  

\section{Conclusion}
\label{conclusion}
We presented \tool, a plug-and-play framework that couples an SVGD-based offline seed generator with online testers via a hazard model, enabling diverse and realistic failure-inducing scenarios. Experiments in CARLA with Apollo~8.0 and Traffic Manager show that \tool outperforms state-of-the-art baselines, improving violation rates by up to 27.68\%, diversity by 9.6\%, and map coverage by 16.78\%, with scenarios rated realistic by human judges.  
\tool’s extensible design allows seamless integration of future online testers and hazard models. A key direction is to enhance both offline seeding and online testing to support higher-fidelity simulation environments, thereby narrowing the simulation–reality gap and enabling safer, more robust ADS deployment.

\section*{Data-Availability Statement}
Our artifact is available at \url{https://doi.org/10.5281/zenodo.19625701} \cite{zenodo_artifact} (mirror: \url{https://github.com/lfeng0722/PtoP}).

\begin{acks}
This work is supported by the Australian Research Council grants
FT240100269 and DP210102447.
\end{acks}
\bibliographystyle{ACM-Reference-Format}
\bibliography{mybibliography}

@article{bertoncello2015ten,
  title={Ten ways autonomous driving could redefine the automotive world},
  author={Bertoncello, Michele and Wee, Dominik},
  journal={McKinsey \& Company},
  volume={6},
  year={2015}
}

@article{barr2014oracle,
  title={The oracle problem in software testing: A survey},
  author={Barr, Earl T and Harman, Mark and McMinn, Phil and Shahbaz, Muzammil and Yoo, Shin},
  journal={IEEE transactions on software engineering},
  volume={41},
  number={5},
  pages={507--525},
  year={2014},
  publisher={IEEE}
}

@inproceedings{li2020av,
  title={Av-fuzzer: Finding safety violations in autonomous driving systems},
  author={Li, Guanpeng and Li, Yiran and Jha, Saurabh and others},
  booktitle={2020 IEEE 31st international symposium on software reliability engineering (ISSRE)}
}

@inproceedings{panichella2015reformulating,
  title={Reformulating branch coverage as a many-objective optimization problem},
  author={Panichella, Annibale and Kifetew, Fitsum Meshesha and Tonella, Paolo},
  booktitle={2015 IEEE 8th international conference on software testing, verification and validation (ICST)},
  pages={1--10},
  year={2015},
  organization={IEEE}
}

@inproceedings{abdessalem2018testing,
  title={Testing autonomous cars for feature interaction failures using many-objective search},
  author={Abdessalem, Raja Ben and Panichella, Annibale and Nejati, Shiva and Briand, Lionel C and Stifter, Thomas},
  booktitle={Proceedings of the 33rd ACM/IEEE International Conference on Automated Software Engineering},
  pages={143--154},
  year={2018}
}

@inproceedings{lu2022rgchaser,
  title={RGChaser: A RL-guided Fuzz and Mutation Testing Framework for Deep Learning Systems},
  author={Lu, Yuteng and Shao, Kaicheng and Sun, Weidi and Sun, Meng},
  booktitle={2022 9th International Conference on Dependable Systems and Their Applications (DSA)},
  pages={12--23},
  year={2022},
  organization={IEEE}
}

@misc{ap,
  title = {{Baidu Apollo team (2017), Apollo: Open Source Autonomous Driving},
  howpublished = {\url{https://github.com/ApolloAuto/apollo}},
  note = {Accessed: 2019-02-11}
}}

@inproceedings{dosovitskiy2017carla,
  title={CARLA: An open urban driving simulator},
  author={Dosovitskiy, Alexey and Ros, German and Codevilla, Felipe and Lopez, Antonio and Koltun, Vladlen},
  booktitle={Conference on robot learning},
  pages={1--16},
  year={2017},
  organization={PMLR}
}

@inproceedings{deng2022scenario,
  title={Scenario-based test reduction and prioritization for multi-module autonomous driving systems},
  author={Deng, Yao and Zheng, Xi and Zhang, Mengshi and Lou, Guannan and Zhang, Tianyi},
  booktitle={Proceedings of the 30th ACM Joint European Software Engineering Conference and Symposium on the Foundations of Software Engineering},
  pages={82--93},
  year={2022}
}

@article{lehman2011abandoning,
  title={Abandoning objectives: Evolution through the search for novelty alone},
  author={Lehman, Joel and Stanley, Kenneth O},
  journal={Evolutionary computation},
  volume={19},
  number={2},
  pages={189--223},
  year={2011},
  publisher={MIT Press}
}

@inproceedings{lehman2008exploiting,
  title={Exploiting open-endedness to solve problems through the search for novelty.},
  author={Lehman, Joel and Stanley, Kenneth O and others},
  booktitle={ALIFE},
  pages={329--336},
  year={2008}
}

@article{zhong2022neural,
  title={Neural network guided evolutionary fuzzing for finding traffic violations of autonomous vehicles},
  author={Zhong, Ziyuan and Kaiser, Gail and Ray, Baishakhi},
  journal={IEEE Transactions on Software Engineering},
  year={2022},
  publisher={IEEE}
}

@inproceedings{tian2022mosat,
  title={MOSAT: finding safety violations of autonomous driving systems using multi-objective genetic algorithm},
  author={Tian, Haoxiang and Jiang, Yan and others},
  booktitle={ESEC/FSE 2022},
  pages={94--106},
  year={2022}
}

@inproceedings{koren2018adaptive,
  title={Adaptive stress testing for autonomous vehicles},
  author={Koren, Mark and Alsaif, Saud and Lee, Ritchie and Kochenderfer, Mykel J},
  booktitle={2018 IEEE Intelligent Vehicles Symposium},
  year={2018},
  organization={IEEE}
}

@article{lu2022learning,
  title={Learning configurations of operating environment of autonomous vehicles to maximize their collisions},
  author={Lu, Chengjie and Shi, Yize and others},
  journal={IEEE Transactions on Software Engineering},
  volume={49},
  number={1},
  pages={384--402},
  year={2022},
  publisher={IEEE}
}

@misc{Apollo,
  author = {ApolloAuto},
  title = {Apollo},
  year = {2024},
publisher = {GitHub},
  journal = {GitHub repository},
  howpublished = {\url{https://github.com/ApolloAuto/apollo}},
  commit = {e373b206a0dc0360af826152132a61c85cab295c}
}

@inproceedings{deb2000fast,
  title={A fast elitist non-dominated sorting genetic algorithm for multi-objective optimization: NSGA-II},
  author={Deb, Kalyanmoy and Agrawal, Samir and Pratap, Amrit and Meyarivan, Tanaka},
  booktitle={Parallel Problem Solving from Nature PPSN VI: 6th International Conference Paris, France, September 18--20, 2000 Proceedings 6},
  pages={849--858},
  year={2000},
  organization={Springer}
}

@article{feng2023dense,
  title={Dense reinforcement learning for safety validation of autonomous vehicles},
  author={Feng, Shuo and Sun, Haowei and Yan, Xintao and others},
  journal={Nature},
  volume={615},
  number={7953},
  year={2023},
  publisher={Nature Publishing Group UK London}
}

@inproceedings{ben2016testing,
  title={Testing advanced driver assistance systems using multi-objective search and neural networks},
  author={Ben Abdessalem, Raja and Nejati, Shiva and Briand, Lionel C and Stifter, Thomas},
  booktitle={Proceedings of the 31st IEEE/ACM international conference on automated software engineering},
  pages={63--74},
  year={2016}
}

@inproceedings{ebadi2021efficient,
  title={Efficient and effective generation of test cases for pedestrian detection-search-based software testing of Baidu Apollo in SVL},
  author={Ebadi, Hamid and Moghadam, Mahshid Helali and others},
  booktitle={2021 IEEE International Conference on Artificial Intelligence Testing (AITest)},
  pages={103--110},
  year={2021},
  organization={IEEE}
}

@article{huai2023sceno,
  title={sceno RITA: Generating Diverse, Fully-Mutable, Test Scenarios for Autonomous Vehicle Planning},
  author={Huai, Yuqi and Almanee, Sumaya and Chen, Yuntianyi and Wu, Xiafa and Chen, Qi Alfred and Garcia, Joshua},
  journal={IEEE Transactions on Software Engineering},
  year={2023},
  publisher={IEEE}
}

@inproceedings{luo2021targeting,
  title={Targeting requirements violations of autonomous driving systems by dynamic evolutionary search},
  author={Luo, Yixing and Zhang, Xiao-Yi and others},
  booktitle={2021 36th IEEE/ACM International Conference on Automated Software Engineering (ASE)},
  pages={279--291},
  year={2021},
  organization={IEEE}
}

@article{lambert2016understanding,
  title={Understanding the fatal tesla accident on autopilot and the nhtsa probe},
  author={Lambert, Fred},
  journal={Electrek, July},
  volume={1},
  pages={1},
  year={2016}
}

@article{liang2023rlaga,
  title={RLaGA: A Reinforcement Learning Augmented Genetic Algorithm For Searching Real and Diverse Marker-Based Landing Violations},
  author={Liang, Linfeng and Deng, Yao and Morton, Kye and Kallinen, Valtteri and James, Alice and Seth, Avishkar and Kuantama, Endrowednes and Mukhopadhyay, Subhas and Han, Richard and Zheng, Xi},
  journal={arXiv preprint arXiv:2310.07378},
  year={2023}
}

@article{deng2023target,
  title={Target: Traffic rule-based test generation for autonomous driving systems},
  author={Deng, Yao and Yao, Jiaohong and Tu, Zhi and Zheng, Xi and Zhang, Mengshi and Zhang, Tianyi},
  journal={arXiv preprint arXiv:2305.06018},
  year={2023}
}

@misc{tesla_autopilot,
  author = {Tesla, Inc.},
  title = {Autopilot},
  year = {2024},
  url = {https://www.tesla.com/en_AU/autopilot},
  note = {Accessed: 2024-11-13}
}

@article{chen2024end,
  title={End-to-end autonomous driving: Challenges and frontiers},
  author={Chen, Li and Wu, Penghao and Chitta, Kashyap and Jaeger, Bernhard and Geiger, Andreas and Li, Hongyang},
  journal={IEEE Transactions on Pattern Analysis and Machine Intelligence},
  year={2024},
  publisher={IEEE}
}

@book{attewell1994crashes,
  title={Crashes resulting in car occupant fatalities: Frontal impacts},
  author={Attewell, Robyn G and Ginpil, Stephen},
  number={CR 138},
  year={1994},
  publisher={Australian Government Pub. Service}
}

@misc{RoadSafetyDataHub2025,
  author       = {{Australian Government Department of Infrastructure, Transport, Regional Development, Communications and the Arts}},
  title        = {Monthly Road Deaths - Road Safety Data Hub},
  year         = {2025},
  url          = {https://datahub.roadsafety.gov.au/progress-reporting/monthly-road-deaths},
  note         = {Accessed: 2025-01-27}
}

@misc{autowarefoundation_autoware,
  author       = {Autoware Foundation},
  title        = {Autoware: Open-Source Software for Autonomous Driving},
  year         = {2025},
  howpublished = {\url{https://github.com/autowarefoundation/autoware}},
  note         = {Accessed: 2025-02-18}
}

@article{cheng2024decictor,
  title={Decictor: Towards Evaluating the Robustness of Decision-Making in Autonomous Driving Systems},
  author={Cheng, Mingfei and Zhou, Yuan and Xie, Xiaofei and Wang, Junjie and Meng, Guozhu and Yang, Kairui},
  journal={arXiv preprint arXiv:2402.18393},
  year={2024}
}

@article{chen2024misconfiguration,
  title={Misconfiguration Software Testing for Failure Emergence in Autonomous Driving Systems},
  author={Chen, Yuntianyi and Huai, Yuqi and Li, Shilong and Hong, Changnam and Garcia, Joshua},
  journal={Proceedings of the ACM on Software Engineering},
  volume={1},
  number={FSE},
  pages={1913--1936},
  year={2024},
  publisher={ACM New York, NY, USA}
}

@misc{Prolific2024,
  author    = {{Prolific}},
  title     = {General citation guidelines},
  year      = {2024},
  howpublished = {Available at \url{https://www.prolific.com}},
  note      = {First released in 2014. Copyright 2024. Located in London, UK. Version: Current month(s) and year(s) of use.}
}

@article{liu2016stein,
  title={Stein variational gradient descent: A general purpose bayesian inference algorithm},
  author={Liu, Qiang and Wang, Dilin},
  journal={Advances in neural information processing systems},
  volume={29},
  year={2016}
}

@inproceedings{rowe2025scenario,
  title={Scenario dreamer: Vectorized latent diffusion for generating driving simulation environments},
  author={Rowe, Luke and Girgis, Roger and Gosselin, Anthony and Paull, Liam and Pal, Christopher and Heide, Felix},
  booktitle={Proceedings of the Computer Vision and Pattern Recognition Conference},
  pages={17207--17218},
  year={2025}
}

@article{sun2023drivescenegen,
  title={Drivescenegen: Generating diverse and realistic driving scenarios from scratch},
  author={Sun, Shuo and Gu, Zekai and Sun, Tianchen and Sun, Jiawei and Yuan, Chengran and Han, Yuhang and Li, Dongen and Ang Jr, Marcelo H},
  journal={arXiv preprint arXiv:2309.14685},
  year={2023}
}

@inproceedings{gambi2019generating,
  title={Generating effective test cases for self-driving cars from police reports},
  author={Gambi, Alessio and Huynh, Tri and Fraser, Gordon},
  booktitle={Proceedings of the 2019 27th ACM Joint Meeting on European Software Engineering Conference and Symposium on the Foundations of Software Engineering},
  pages={257--267},
  year={2019}
}

@inproceedings{hauer2020clustering,
  title={Clustering traffic scenarios using mental models as little as possible},
  author={Hauer, Florian and Gerostathopoulos, Ilias and Schmidt, Tabea and Pretschner, Alexander},
  booktitle={2020 IEEE Intelligent Vehicles Symposium (IV)},
  pages={1007--1012},
  year={2020},
  organization={IEEE}
}

@inproceedings{de2017assessment,
  title={Assessment of automated driving systems using real-life scenarios},
  author={De Gelder, Erwin and Paardekooper, Jan-Pieter},
  booktitle={2017 ieee intelligent vehicles symposium (iv)},
  pages={589--594},
  year={2017},
  organization={IEEE}
}

@article{li2024generalization,
  title={Generalization of cut-in pre-crash scenarios for autonomous vehicles based on accident data},
  author={Li, Pingfei and Zhu, Xinyu and Ren, Yao and Tan, Zhengping and Hu, Wenhao and Zhang, You and Xu, Chang},
  journal={Scientific reports},
  volume={14},
  number={1},
  pages={17664},
  year={2024},
  publisher={Nature Publishing Group UK London}
}

@inproceedings{kluck2018using,
  title={Using ontologies for test suites generation for automated and autonomous driving functions},
  author={Kl{\"u}ck, Florian and Li, Yihao and Nica, Mihai and Tao, Jianbo and Wotawa, Franz},
  booktitle={2018 IEEE International symposium on software reliability engineering workshops (ISSREW)},
  pages={118--123},
  year={2018},
  organization={IEEE}
}

@inproceedings{birkemeyer2022feature,
  title={Feature-Interaction Sampling for Scenario-based Testing of Advanced Driver Assistance Systems},
  author={Birkemeyer, Lukas and Pett, Tobias and Vogelsang, Andreas and Seidl, Christoph and Schaefer, Ina},
  booktitle={Proceedings of the 16th International Working Conference on Variability Modelling of Software-Intensive Systems},
  pages={1--10},
  year={2022}
}

@inproceedings{huai2023doppelganger,
  title={Doppelg{\"a}nger test generation for revealing bugs in autonomous driving software},
  author={Huai, Yuqi and Chen, Yuntianyi and Almanee, Sumaya and Ngo, Tuan and Liao, Xiang and Wan, Ziwen and Chen, Qi Alfred and Garcia, Joshua},
  booktitle={2023 IEEE/ACM 45th International Conference on Software Engineering (ICSE)},
  pages={2591--2603},
  year={2023},
  organization={IEEE}
}

@inproceedings{hanselmann2022king,
  title={King: Generating safety-critical driving scenarios for robust imitation via kinematics gradients},
  author={Hanselmann, Niklas and Renz, Katrin and Chitta, Kashyap and Bhattacharyya, Apratim and Geiger, Andreas},
  booktitle={European Conference on Computer Vision},
  pages={335--352},
  year={2022},
  organization={Springer}
}

@inproceedings{chen2004adaptive,
  title={Adaptive random testing},
  author={Chen, Tsong Yueh and Leung, Hing and Mak, Ieng Kei},
  booktitle={Annual Asian Computing Science Conference},
  pages={320--329},
  year={2004},
  organization={Springer}
}

@article{mnih2013playing,
  title={Playing atari with deep reinforcement learning},
  author={Mnih, Volodymyr and Kavukcuoglu, Koray and Silver, David and Graves, Alex and Antonoglou, Ioannis and Wierstra, Daan and Riedmiller, Martin},
  journal={arXiv preprint arXiv:1312.5602},
  year={2013}
}

@misc{zenodo_artifact,
  author       = {Linfeng Liang and Xiao Cheng and Tsong Yueh Chen and Xi Zheng},
  title        = {Artifact for: From Particles to Perils: SVGD-Based Hazardous Scenario Generation for Autonomous Driving Systems Testing},
  year         = {2025},
  publisher    = {Zenodo},
  doi          = {10.5281/zenodo.19625701},
  url          = {https://doi.org/10.5281/zenodo.19625701}
}
%%
%% The next two lines define the bibliography style to be used, and
%% the bibliography file.

%%
%% If your work has an appendix, this is the place to put it.
% \appendix

% \section{Research Methods}

% \subsection{Part One}

% Lorem ipsum dolor sit amet, consectetur adipiscing elit. Morbi
% malesuada, quam in pulvinar varius, metus nunc fermentum urna, id
% sollicitudin purus odio sit amet enim. Aliquam ullamcorper eu ipsum
% vel mollis. Curabitur quis dictum nisl. Phasellus vel semper risus, et
% lacinia dolor. Integer ultricies commodo sem nec semper.

% \subsection{Part Two}

% Etiam commodo feugiat nisl pulvinar pellentesque. Etiam auctor sodales
% ligula, non varius nibh pulvinar semper. Suspendisse nec lectus non
% ipsum convallis congue hendrerit vitae sapien. Donec at laoreet
% eros. Vivamus non purus placerat, scelerisque diam eu, cursus
% ante. Etiam aliquam tortor auctor efficitur mattis.

% \section{Online Resources}

% Nam id fermentum dui. Suspendisse sagittis tortor a nulla mollis, in
% pulvinar ex pretium. Sed interdum orci quis metus euismod, et sagittis
% enim maximus. Vestibulum gravida massa ut felis suscipit
% congue. Quisque mattis elit a risus ultrices commodo venenatis eget
% dui. Etiam sagittis eleifend elementum.

% Nam interdum magna at lectus dignissim, ac dignissim lorem
% rhoncus. Maecenas eu arcu ac neque placerat aliquam. Nunc pulvinar
% massa et mattis lacinia.

\end{document}